\documentclass[%
 reprint,
superscriptaddress,
 amsmath,amssymb,
 aps,
prm,
]{revtex4-1}

\usepackage{graphicx}
\usepackage{epstopdf}
\usepackage{dcolumn}
\usepackage{bm}
\usepackage{gensymb}
\usepackage{upgreek}
\usepackage{hyperref}
\usepackage{tabularx}
\usepackage{chngcntr}

\preprint{APS/123-QED}

\begin{document}

\title{Real-space orbital tiling approach for the design of novel superconductors}
\author{Gregory Bassen}
\email[]{gbassen1@jhu.edu}
\affiliation{Institute for Quantum Matter, William H. Miller III Department of Physics and Astronomy, Johns Hopkins University, 3400 N. Charles Street, Baltimore, MD 21218, United States of America}
\affiliation{Department of Chemistry, Johns Hopkins University, 3400 N. Charles Street, Baltimore, MD 21218, United States of America}
\affiliation{Department of Materials Science and Engineering, Johns Hopkins University, 3400 N. Charles Street, Baltimore, MD 21218, United States of America}
\author{Wyatt Bunstine}
\affiliation{Institute for Quantum Matter, William H. Miller III Department of Physics and Astronomy, Johns Hopkins University, 3400 N. Charles Street, Baltimore, MD 21218, United States of America}
\affiliation{Department of Chemistry, Johns Hopkins University, 3400 N. Charles Street, Baltimore, MD 21218, United States of America}
\affiliation{Department of Materials Science and Engineering, Johns Hopkins University, 3400 N. Charles Street, Baltimore, MD 21218, United States of America}
\author{Rebecca Han}
\affiliation{Institute for Quantum Matter, William H. Miller III Department of Physics and Astronomy, Johns Hopkins University, 3400 N. Charles Street, Baltimore, MD 21218, United States of America}
\affiliation{Department of Chemistry, Johns Hopkins University, 3400 N. Charles Street, Baltimore, MD 21218, United States of America}
\affiliation{Department of Materials Science and Engineering, Johns Hopkins University, 3400 N. Charles Street, Baltimore, MD 21218, United States of America}
\author{Ragy Ebeid}
\affiliation{Institute for Quantum Matter, William H. Miller III Department of Physics and Astronomy, Johns Hopkins University, 3400 N. Charles Street, Baltimore, MD 21218, United States of America}
\affiliation{Department of Chemistry, Johns Hopkins University, 3400 N. Charles Street, Baltimore, MD 21218, United States of America}
\affiliation{Department of Materials Science and Engineering, Johns Hopkins University, 3400 N. Charles Street, Baltimore, MD 21218, United States of America}
\author{Eli Zoghlin}
\affiliation{Institute for Quantum Matter, William H. Miller III Department of Physics and Astronomy, Johns Hopkins University, 3400 N. Charles Street, Baltimore, MD 21218, United States of America}
\affiliation{Department of Chemistry, Johns Hopkins University, 3400 N. Charles Street, Baltimore, MD 21218, United States of America}
\affiliation{Department of Materials Science and Engineering, Johns Hopkins University, 3400 N. Charles Street, Baltimore, MD 21218, United States of America}
\author{Tyrel M. McQueen}
\email[]{mcqueen@jhu.edu}
\affiliation{Institute for Quantum Matter, William H. Miller III Department of Physics and Astronomy, Johns Hopkins University, 3400 N. Charles Street, Baltimore, MD 21218, United States of America}
\affiliation{Department of Chemistry, Johns Hopkins University, 3400 N. Charles Street, Baltimore, MD 21218, United States of America}
\affiliation{Department of Materials Science and Engineering, Johns Hopkins University, 3400 N. Charles Street, Baltimore, MD 21218, United States of America}

\date{\today}

\begin{abstract}
Despite substantial advances in the field, we still lack a predictive framework capable of guiding the discovery of new families of superconductors. While momentum-space approaches have advanced the microscopic understanding of superconductivity, they offer limited guidance for materials design based on atomic building blocks. Here, we propose a real-space framework which conceptualizes Cooper pairs as confined standing waves resulting from coherent tilings of atomic orbitals. We call this model the Real-space Orbital Superconducting Pathway (ROSP). Using a tight-binding toy model, we show that the energetics of electron pairing depend on the configuration and overlap of real-space orbitals, which motivates \textit{a priori} design of superconducting families from orbital tiling. We connect the ROSP model to Roald Hoffmann’s isolobal analogy to classify families of superconductors based on shared orbital tilings, rather than structure or electron count. As an example, we suggest that superconductivity in La$_{3}$Ni$_{2}$O$_{7}$ and LaNiO$_{2}$, despite differing structures and electron counts, may arise from a common ROSP. We introduce a new notation to classify two-dimensional square-net ROSPs and further propose several new families of superconductors on the anti-cuprate lattice. This framework provides a new model for predicting and designing families of high-T$_c$ superconductors from real-space orbital architecture, even without microscopic knowledge of the attractive pairing interaction.

\end{abstract}

\maketitle

\section{Introduction}

A definitive microscopic model for high-temperature superconductivity remains one of the largest unsolved mysteries in condensed matter physics. Furthermore, the occurrence of high-$T_c$ superconductivity among families of unconventional superconductors has inextricably linked this effort with the desire to discover ever higher $T_c$ superconductors for technological applications. The ongoing synthesis and discovery of new superconductors have played a critical role in stimulating advancements in theoretical understanding and providing empirical clues to the underlying mechanisms of unconventional superconductivity. Ideally, materials discovery and theory operate in a feedback loop, where theoretical predictions guide synthetic efforts, and experimental outcomes, in turn, validate or challenge theoretical models. In this context, the ideal theory would not only explain the mechanism of unconventional superconductivity, but also point toward materials where $T_c$ or other relevant properties, such as critical current and irreversibility field, can be maximized. However, predictive discovery has not proven to be straightforward. While a wide array of theories have been proposed to explain the mechanism underlying unconventional superconductivity—such as the ground-state doped square lattice Hubbard model, non-linear electron-phonon coupling, and others \cite{senthil_coherence_2009,yin_correlation-enhanced_2013,zheng_ground-state_2016,babadi_theory_2017}—none have yet succeeded in predicting new families of high-T$_c$ superconductors. In practice, the discovery of new unconventional superconductors has been driven by a combination of experimental heuristics and serendipity. 

For example, the seminal discovery by Bednorz and M{\"u}ller in 1986 of high-temperature superconductivity in the layered perovskite La$_{2-x}$Ba$_{x}$CuO$_{4}$ (LBCO) was guided by a simple heuristic: selecting Jahn-Teller active ions, such as Ni and Cu in perovskite-type oxides, in order to enhance electron-phonon coupling, as inspired by microscopic considerations from the then well-established Bardeen-Cooper-Schrieffer (BCS) theory \cite{bednorz_possible_1986,bednorz_perovskite-type_1988}. Although the superconducting state of LBCO is not fully explained within the BCS framework, the heuristic nonetheless provided a critical avenue for translating theoretical concepts into experimental discovery, even in the absence of a complete theoretical framework.

Bednorz and M{\"u}ller were specifically interested in exploring structures with perovskite motifs due to previous reports of superconductivity in SrTiO$_{3}$ (T$c$ $\approx$  0.25 K, 1964) \cite{schooley_superconductivity_1964} and BaPb$_{1-x}$Bi$_{x}$O$_{3}$ (T$c$ $\approx$  13 K, 1975) \cite{sleight_high-temperature_1975}.  Guided by these precedents, they treated the perovskite lattice as a platform for chemical exploration within their emerging heuristic. A similar drive to explore perovskites led Cava et al., early in the cuprate era (1988), to investigate a related bismuthate, Ba$_{1-x}$K$_{x}$BiO$_{3}$, ultimately discovering superconductivity at T$_c$ $\approx$  30 K \cite{cava_superconductivity_1988}. 

 It did not take long for the community to discover a wide variety of cuprate compounds with progressively higher T$_c$'s. This was made possible by recognizing the central role of the CuO$_2$ layer and employing chemical design strategies to tune the electronic structure by modifying the interlayer distance, the Cu charge state within the layers, and the local coordination environment around the Cu ions \cite{park_structures_1995}. This strategy — targeting structural motifs known to support superconductivity and systematically expanding the chemical space around them — became the dominant approach of the solid-state community to explore families of superconductors. The success of Bednorz and Müller's heuristic also inspired broader discoveries beyond layered perovskites, particularly toward materials thought to similarly maximize electron-phonon coupling. A notable example was the synthesis of the first icosahedral C$_{60}$ buckminsterfullerene molecule in 1985 \cite{kroto_c60_1985}, which led Hebard et al. to intercalate alkali metals into C$_{60}$, culminating in the discovery of superconductivity at T$_c$ $\approx$ 18 K in K${_3}$C${_{60}}$ in 1991 \cite{hebard_superconductivity_1991}. Yet even as heuristic-driven searches expanded, serendipitous discoveries continued to play a central role. In 1996, superconductivity was unexpectedly found in lithium-intercalated $\beta$-ZrNCl—a wide-band-gap insulator with a 3 eV gap—at T$_c$ $\approx$  12 K \cite{yamanaka_new_1996}. Shortly thereafter, superconductivity at an even higher T$_c$ of 25.5 K was discovered in the isostructural Li$_x$HfNCl \cite{yamanaka_superconductivity_1998}. 

These examples underscore that, while experimental heuristics can guide exploration within known paradigms, they are not sufficient to foresee the emergence of entirely new families of superconductors. Moreover, although a vast theoretical literature on unconventional superconductivity developed in parallel with these experimental successes, it did not directly guide the discovery of the next major family of unconventional superconductors. Instead, the iron-based superconductors emerged serendipitously in 2006 during a search for transparent oxide conductors, with the discovery of superconductivity in LaFePO (T$_c$ $\approx$  4 K) \cite{ma_progress_2012,kamihara_iron-based_2006}. As with the cuprates, once the essential structural motif — a layer of edge-sharing, tetrahedrally coordinated Fe ions — was identified, rapid optimization within the phase space quickly followed. The iron-based family was subsequently expanded using similar chemical design strategies pioneered during the cuprate era, namely fine-tuning the layer separation, the Fe charge state, and the local coordination environment.

The difficulty of discovering superconductors from theoretical considerations alone may stem, in part, from the non-intuitive nature of mapping momentum-space concepts onto real-space structures. There has long been an effort to bridge the momentum-space thinking of condensed matter physics with the real-space, orbital-based thinking characteristic of molecular chemistry, particularly as pioneered by Roald Hoffmann \cite{hoffmann_how_1987, skorupskii_designing_2024, jovanovic_simple_2022, xiao_conduction_2014}. Nevertheless, a clear divide between these two perspectives remains. This is evident in field-specific approaches to electrical conductivity: for example, in the metal-organic framework (MOF) community, conductivity is discussed in real-space, and described as charge transfer through-bond, through-space, and via extended conjugation pathways \cite{xie_electrically_2020}. These pathways arise from a molecular and orbital understanding of the constituent frameworks and how charge moves through them. By contrast, in the condensed matter physics community, conductivity is commonly conceptualized through the free-electron model, which treats electron motion statistically as a non-interacting Fermi gas moving across a featureless background potential.

More recently, however, there has been growing interest in mapping conduction phenomena in real-space. For example, space-projected conductivity (SPC) has been developed to visualize electronic probability densities through local structural units in materials such as metal composites, conducting-bridge random-access memory (CBRAM), phase-change memory (PCM), and atomic switch devices \cite{subedi_modeling_2021, xiao_conduction_2014, muriel_introducing_2024, nepal_physical_2024}. SPC methods provide real-space insights into where and how conduction occurs at the atomic scale \cite{prasai_spatial_2018, subedi_spaceprojected_2021}. Similarly, in high-T$_c$ high-pressure hydride superconductors, real-space visualization approaches have been employed, using a "networking value" based on electronic localization functions and isosurface analyses to characterize the extent of electronic delocalization around atomic sites \cite{muriel_introducing_2024, belli_strong_2021}. Research along these lines may be indicative of real-space conceptualizations of electronic behavior being embraced in condensed matter physics.

Early efforts to apply real-space thinking to superconductivity anticipated some of the more recent trends toward orbital- and structure-based descriptions of electronic behavior. Much like conductivity in MOFs, superconductivity has also been approached from a real-space, molecular-orbital perspective. In 1975, Krebs proposed an analogy between the $p\pi$ orbitals of aromatic benzene and the electronic wavefunctions of extended solids, suggesting that superconductivity requires a bonding network capable of supporting a molecular-like wavefunction uninterrupted by a plane or conical nodal surface along at least one spatial direction \cite{krebs_superconductivity_1975}. Krebs further argued that the corresponding electronic band must be partially filled for superconductivity to be realized. Extending this line of thought, in 1983 Johnson et al.  mapped the electronic networks of intermetallic superconductors by analyzing the highest occupied molecular orbitals (HOMOs), which we will herein refer to as \textit{frontier orbitals}, of clustered structural units \cite{johnson_molecular-orbital_1983}.

The influence of orbital character on superconductivity has been increasingly studied in recent years. Theoretical models have explored orbital-selective Mott transitions \cite{koga_orbital-selective_2004,de_medici_orbital-selective_2009} and the spontaneous emergence of orbital-selective superconductivity from such Mott states \cite{ishigaki_spontaneously_2018}. Similarly, orbital-selective Cooper pairing has been proposed within a two-band model for nickelates \cite{adhikary_orbital-selective_2020}. Experimentally, the iron-based superconductors A$_x$Fe$_{2-y}$Se$_2$ (A = K, Rb) have been shown to exhibit an orbital-selective Mott phase and a metal-insulator transition above T$_c$ \cite{wang_orbital-selective_2014,yi_observation_2013}, while Sr$_2$RuO$_4$ shows evidence of orbital dependence and ordering near its superconducting transition \cite{agterberg_orbital_1997}. Beyond these examples, Das et al. demonstrated that Weyl semimetals and superconductors can be designed through an orbital-selective superlattice of layered even- and odd-parity orbitals \cite{das_weyl_2013}. Together, these band and orbital studies establish that superconductivity is strongly influenced by orbital contributions. 

Viewed in this broader context, the discovery of new superconductors has consistently followed from the chemical exploration of new structures, particularly with the development of modern chemical design strategies \cite{drozdov_superconductivity_2019, xie_endohedral_2015, canfield_new_2020, whoriskey_amalgams_2024}, and is typically driven by a blend of heuristics and serendipity. However, despite these historical successes, the theoretical insights gained from these discoveries have not yet translated into a predictive framework for identifying new superconducting families. In the absence of a complete microscopic theory of unconventional superconductivity, there remains a need for improved heuristics that can guide synthetic exploration across promising chemical phase spaces. It is evident from the two families of high-T$_c$ superconductors that certain underlying structural and electronic requirements must be satisfied for superconductivity to emerge. Moreover, although orbital character is increasingly recognized as critical to superconductivity, how real-space orbital configurations can be applied for the design of new superconductors has yet to be clarified. 

The goal of this perspective is to offer a framework based on real-space orbital tiling for discovering new families of high-T$_c$ superconductors. We begin by introducing a simple orbital toy model that illustrates how the energetics of an effective attractive electron-electron interaction depend on the tiling of atomic orbitals. We propose that superconductivity can be viewed as the formation of Cooper pair standing waves along a discrete real-space tiling of orbitals, a model we refer to as the Real-space orbital superconducting pathway (ROSP). We then outline chemical design strategies to realize ROSP motifs within known structure types and propose candidate materials exhibiting these features. In addition, we propose new families of superconductors on the anti-cuprate lattice as corollaries to known perovskite derived families.  Finally, we illustrate how orbital tiling principles can be used to design new superconducting structures from the ground up, translating theoretical ROSPs into realizable materials.

\begin{figure}
    \centering
    \includegraphics[height = 3.33in, width = 3.33in, keepaspectratio]{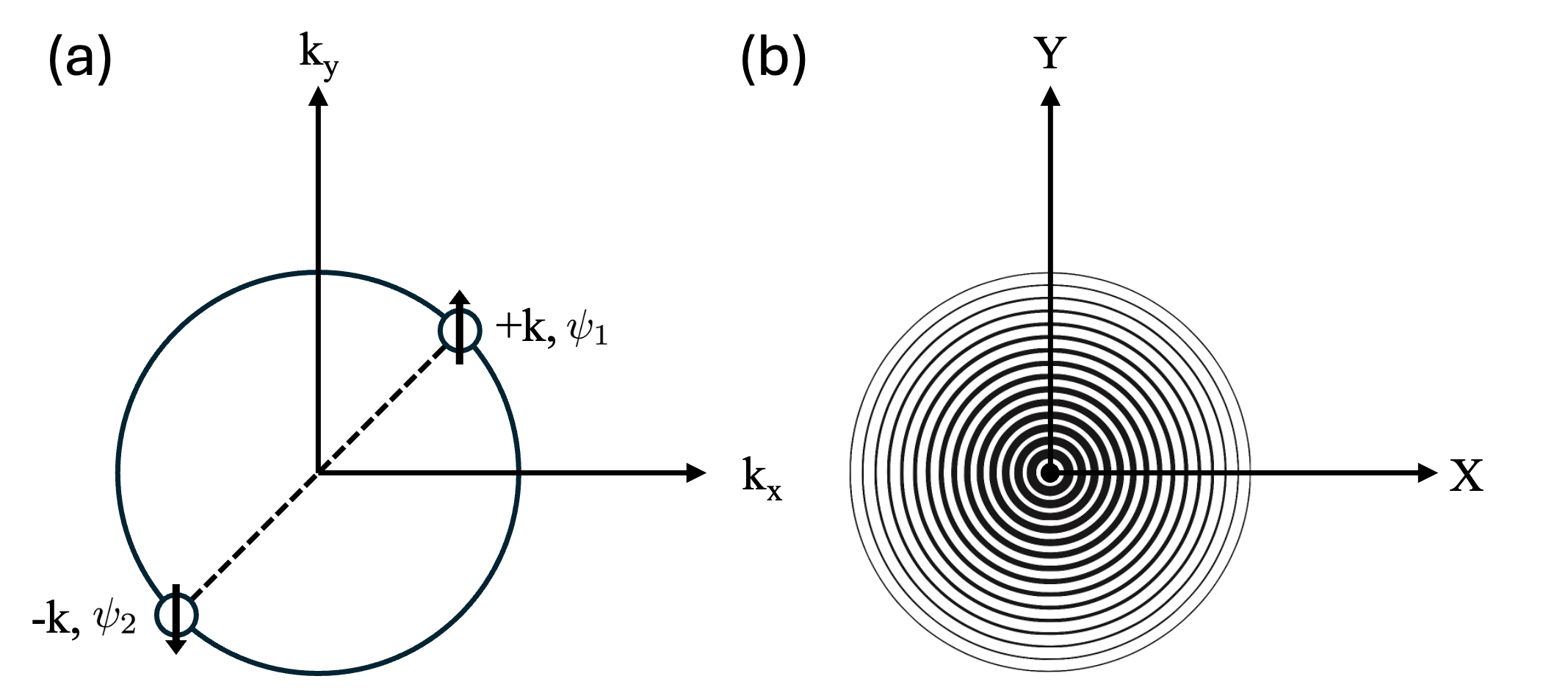}
    \caption{a) Schematic illustration, in momentum space, of BCS Cooper pairing between electrons with opposite spin and momentum on a two-dimensional Fermi surface. b) Schematic of the real-space probability density for the $s$-wave symmetric Cooper pair shown in (a), where darker concentric regions indicate higher probability of finding the paired electrons.}
    \label{fig:Schematic}
\end{figure}

\section{Results and Discussion}

\subsection{Real-space Orbital Superconducting Pathway}
As previously described, electrons are commonly modeled in momentum space, such as through Bloch functions. Accordingly, superconductivity is largely studied in momentum space, and the Fermi surface is primarily used to understand electronic behavior such as pairing mechanisms and the nature of the superconducting gap. Although much is still unknown about superconductors, there are several important theoretical characteristics that are foundational for developing real-space intuition. The Ginzburg-Laundau theory posits that the superconducting state can be described phenomenologically by a complex order parameter, or, in other words, as a single coherent wavefunction below T$_{c}$ \cite{ginzburg_theory_2009}, as shown in equation \ref{eq:wave_function_representation}.  
\begin{equation}
    \Psi(\mathbf{r}) = |\psi(\mathbf{r})| e^{i \theta(\mathbf{r})}
    \label{eq:wave_function_representation}
\end{equation}

One may begin to develop a local real-space model of superconductivity using the microscopic interpretation of BCS theory, where superconducting electrons form Cooper pairs due to an attractive force, and condense into a single quantum state as bosonic quasiparticles \cite{Cooper1956,Bardeen1957}. Figure \ref{fig:Schematic}(a) shows a schematic of conventional BCS Cooper pairing in momentum space on a two-dimensional Fermi surface at $E=E_{{f}}$. Figure \ref{fig:Schematic}(b) shows the probability density of finding this Cooper pair in real-space in the absence of a lattice.

To make a qualitative toy model, we apply the tight-binding model for electrons in a two-dimensional square lattice with lattice constant $a$. The canonical tight binding Hamiltonian is given by: 
\begin{equation}
  H_{tb}(\mathbf{r}) = H_{\text{at}}(\mathbf{r}) + \sum_{\mathbf{R}_n \neq 0} V(\mathbf{r} - \mathbf{R}_n) ,
  \label{eq:H}
\end{equation}where $H_{\text{at}}(\mathbf{r})$ is the Hamiltonian for an isolated atom and $V(\mathbf{r} - \mathbf{R}_n)$ is the potential from an atom at site n, with corresponding lattice vector $\mathbf{R}_n$. As we are exploring this model through the lens of BCS theory, we assume there is an effective attractive potential between electrons. A completely rigorous approach would require this term to be added to the Hamiltonian from the outset. In practice, this term is small compared to the atomic potentials, so we will neglect it when constructing the initial wavefunctions. In the canonical approach to the tight binding model, one constructs wavefunctions based on atomic orbitals, $\varphi_m(\mathbf{r}-\mathbf{R}_n)$. For $N$ sites at the $m$-th atomic energy level, the one-electron wavefunctions are written as a linear combination of atomic orbitals (LCAO), and we will first use $s$-orbitals as the basis:

\begin{equation}
    \psi_m(\mathbf{r}) \approx \frac{1}{\sqrt{N}}\sum_{\mathbf{R}_n} e^{i\mathbf{k}\cdot\mathbf{R}_n}\varphi_m(\mathbf{r}-\mathbf{R}_n)
    \label{eq:LCOA}
\end{equation}

In this model then, $\psi_{1}$ and $\psi_{2}$ from Figure \ref{fig:Schematic}(a) are solutions to $H_{tb}$ with an energy $\epsilon_0$, the Fermi energy. Each of these states has both a spin-up and spin-down copy.

As all solutions of the same energy are degenerate, we are allowed to take linear combinations to form new single-electron states, i.e. $\psi(r)=a \psi{_1}(r)+b \psi{_2}(r)$ for arbitrarily complex numbers $a$ and $b$, with $|a^2|+|b^2|=1$. For the purposes of building a real-space model of superconductivity, it is useful to take equally weighted combinations of $\psi_{1}$ and $\psi_{2}$, i.e. $\psi(\mathbf{r}) = \frac{1}{\sqrt{2}}\left(e^{i\theta_1}\psi_1(\mathbf{r}) + e^{i\theta_2}\psi_2(\mathbf{r})\right)$ or, equivalently:
 \textbf{\begin{figure}
    \centering
    \includegraphics[height = 3.33in, width = 3.33in, keepaspectratio]{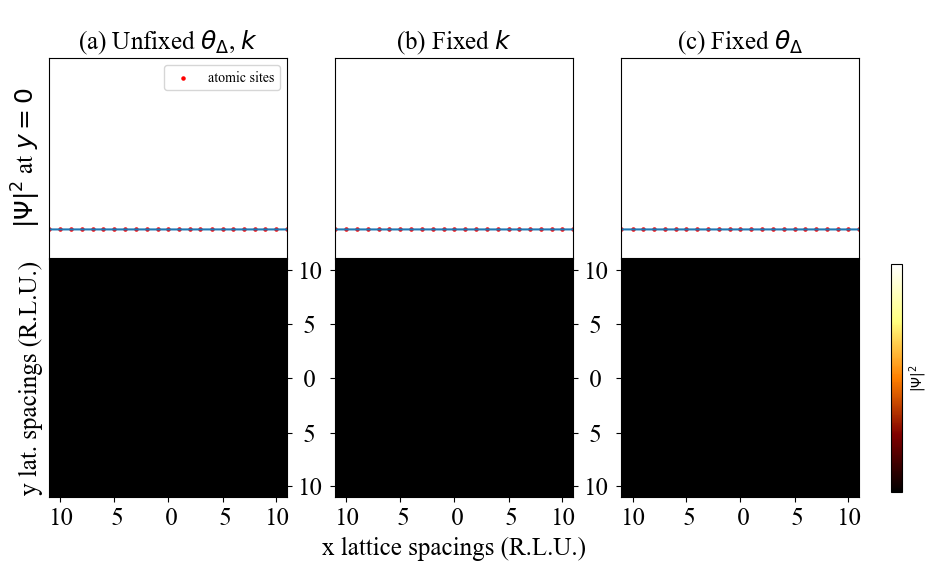}
    \caption{$|\Psi|^2$ linecuts and heatmaps for trivial cases without electron pairing: a) Snapshot with unfixed phase relations with randomly paired Fermi surface states $\psi_1$ and $\psi_2$. b) Snapshot with unfixed phase relations but with $\psi_1$ and $\psi_2$ Fermi surface states with equal and opposite momenta. c) Snapshot with $\psi_1$ and $\psi_2$ Fermi surface states with equal and opposite momenta, and fixed $\theta_\Delta$ within pairs, but unfixed phases between pairs. These plots demonstrate that there is no standing wave in any case where some of the phases remain random.}
    \label{fig:Uniform_Calculation}
\end{figure}}
\begin{equation}
   \psi(\mathbf{r}) = \frac{e^{i\theta_1}}{\sqrt{2}}\left(\psi_1(\mathbf{r}) + e^{i\theta_\Delta}\psi_2(\mathbf{r})\right) ,
    \label{eq:single_particle}
\end{equation} where $\theta_\Delta$ = $\theta_2$ - $\theta_1$, and $\psi_{1}$ and $\psi_{2}$ are chosen to have opposite spin. In the normal state, i.e. in the absence of additional interactions, these new wavefunctions $\psi(r)$ have the same energy for all $\theta_1$ and $\theta_\Delta$ (and for all $\psi_{1}$ and $\psi_{2}$ on the Fermi surface). If we sum up all such single particle states, corresponding to all randomly selected, distinct pairs of $\psi_{1}$ and $\psi_{2}$ with randomly assigned $\theta_1$ and $\theta_\Delta$, we get a real picture of the system at an instant. As shown in Figure \ref{fig:Uniform_Calculation}(a), this results in a featureless map, as expected since there is no phase coherence between electron states. Fixing the choice of pairs to have a precise momentum relation ($\psi_{1}= +\mathbf{k}$ and $\psi_{2}= -\mathbf{k}$), the sum with randomly assigned $\theta_1$ and $\theta_\Delta$ is also featureless, Figure \ref{fig:Uniform_Calculation}(b). Even fixing the choice of pairs to have a precise momentum relation ($\psi_{1}= +\mathbf{k}$ and $\psi_{2}= -\mathbf{k}$), and fixed $\theta_\Delta$, the sum with randomly assigned $\theta_1$ is still featureless, Figure \ref{fig:Uniform_Calculation}(c). This last case corresponds to a system with formed Cooper pairs, but without phase coherence between pairs.

However, something different happens if both the momentum and phase relationships within and between pairs are fixed. For simplicity, assume $k_{pair} = \mathbf{k}{_1}+\mathbf{k}{_2}=0$, i.e. $\psi_{1}= +\mathbf{k}$ and $\psi_{2}= -\mathbf{k}$. Suppose furthermore that we set $\theta_1=fixed$ and $\theta_\Delta=fixed$. Without loss of generality, we will take $\theta_1=0$ yielding:

\begin{equation}
   \psi(\mathbf{r}) = \frac{1}{\sqrt{2}}\left(\psi_{{+\mathbf{k}}}(\mathbf{r}) + e^{i\theta_\Delta}\psi_{{-\mathbf{k}}}(\mathbf{r})\right) .
    \label{eq:single_particle_pair}
\end{equation}

 Summing up all such pairs with $\theta_\Delta=fixed$, something different happens, as shown in Figure \ref{fig:Calculation}. We now see a net feature, a standing wave, that is strongest at $\theta_\Delta=0$, and vanishes at $\theta_\Delta=\pi$. This scenario sounds familiar: fixing momentum and $\theta_\Delta$ is reminiscent of creating a Cooper pair, and fixing phases $\theta_1$ between pairs is reminiscent of forming a superconducting state. In this limit, we argue that this is exactly what Equation \ref{eq:single_particle_pair} describes -- the "single particle picture," wave function that makes a superconducting state. We note at this point that this single particle wavefunction is \underline{not} a Cooper pair. A traditional Cooper pair (for a singlet superconductor with $k_{pair}=0$) is a symmetric product state for the spatial component, given for general wave function $\tilde{\psi}$ as: 
\begin{equation}
\begin{split}
\begin{aligned} 
    \Psi_{Spatial}(\mathbf{r_{1}},\mathbf{r_{2}}) =\\
   \frac{1}{\sqrt{2}}\left(\tilde{\psi}_{{+\mathbf{k}}}(\mathbf{r_{1}}) \tilde{\psi}_{{-\mathbf{k}}}(\mathbf{r_{2}})+\tilde{\psi}_{{+\mathbf{k}}}(\mathbf{r_{2}}) \tilde{\psi}_{{-\mathbf{k}}}(\mathbf{r_{1}})\right) .
\end{aligned}
\end{split}
\label{eq:Psi_CooperPair}
\end{equation}

This is made overall antisymmetric by the spin component (not shown). It should be immediately clear that a product state of two electrons, each described by \ref{eq:single_particle_pair}, is also symmetric: 
\begin{equation}
\begin{split}
\begin{aligned} 
 \Psi_{Spatial}(\mathbf{r_{1}},\mathbf{r_{2}}) = \psi(\mathbf{r_{1}})\psi(\mathbf{r_{2}}) = \psi(\mathbf{r_{2}})\psi(\mathbf{r_{1}}) = \\
   \frac{1}{2}[\psi_{{+\mathbf{k}}}(\mathbf{r_{1}}) \psi_{{+\mathbf{k}}}(\mathbf{r_{2}})
   + e^{i\theta_\Delta}\psi_{{+\mathbf{k}}}(\mathbf{r_{1}}) \psi_{{-\mathbf{k}}}(\mathbf{r_{2}})\\\
   + e^{i\theta_\Delta}\psi_{{-\mathbf{k}}}(\mathbf{r_{1}}) \psi_{{+\mathbf{k}}}(\mathbf{r_{2}})+e^{2i\theta_\Delta}\psi_{{-\mathbf{k}}}(\mathbf{r_{1}}) \psi_{{-\mathbf{k}}}(\mathbf{r_{2}})]
\end{aligned}
\end{split}
\label{eq:Psi_CooperPair_expanded}
\end{equation} Note that in the presence of inversion symmetry [$\psi_{{+\mathbf{k}}}(\mathbf{r})=\psi_{{-\mathbf{k}}}(\mathbf{r})$], and equation \ref{eq:Psi_CooperPair_expanded} becomes identical, up to a phase, with equation \ref{eq:Psi_CooperPair} when $\theta_\Delta=\frac{\pi}{2}$, and goes to zero (i.e. no stable paired state) when $\theta_\Delta={\pi}$. So our carefully chosen linear combination of Fermi surface states, Equation \ref{eq:single_particle_pair}, is a "one electron" representation of the spatially symmetric state found in a superconductor.

\textbf{\begin{figure}
    \centering
    \includegraphics[height = 3.33in, width = 3.33in, keepaspectratio]{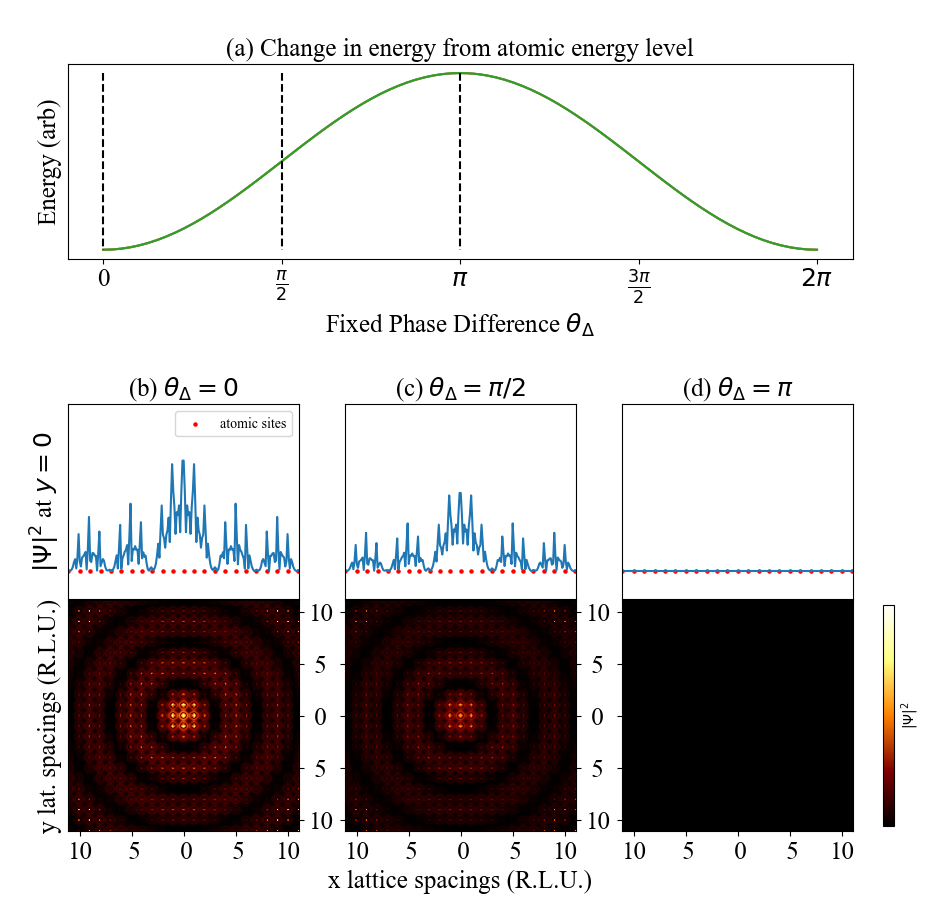}
    \caption{a) Energy of a "single particle" superconducting wavefunction as a function of the fixed phase difference, $\theta_\Delta$, calculated using a tight-binding model of $s$ orbitals. Probability density of finding such a state along atomic sites with $\theta_\Delta$ =  b) 0, c) $\pi/2$, d) $\pi$, where brighter coloration corresponds to higher probability}
    \label{fig:Calculation}
\end{figure}}

Now that we have a "one electron" representation of a superconducting state we should ask what kind of interactions would break the energy degeneracy of Equation \ref{eq:single_particle_pair} as a function of $\theta_\Delta$, i.e. result in selection of some choices of phase over others. Importantly, even in the absence of a specific microscopic interaction, general statements can still be made. Consider some additional interaction $H_{perturb}$ that acts on our pair of one-electron states, $\psi_{1}$ and $\psi_{2}$. The energies of equation \ref{eq:single_particle_pair} become dependent on $\theta_\Delta$ when $\langle \psi_{+\mathbf{k}}(\mathbf{r}) | H_{{perturb}} | \psi_{-\mathbf{k}}(\mathbf{r}) \rangle = \epsilon_p$ are nonzero. This can be represented in matrix form as:

\begin{equation}
\begin{split}
\begin{aligned} 
    \hat{H}_{perturb}=
    \begin{pmatrix}
        \epsilon_0 & \epsilon_p \\
        \epsilon_p & \epsilon_0
    \end{pmatrix}
\end{aligned}
\end{split}
\end{equation}

The solutions of this Hamiltonian are $\frac{1}{\sqrt{2}}[\psi_{{+\mathbf{k}}}(\mathbf{r})+\psi_{{-\mathbf{k}}}(\mathbf{r})]$ with $ \epsilon=\epsilon_0 + \epsilon_p$, or$\frac{1}{\sqrt{2}}[\psi_{{+\mathbf{k}}}(\mathbf{r})-\psi_{{-\mathbf{k}}}(\mathbf{r})]$ with $ \epsilon=\epsilon_0 - \epsilon_p$. If this additional interaction is repulsive (i.e. $ \epsilon_p > 0$), then the latter solution (which corresponds to $\theta_\Delta=\pi$) is lower in energy and there is no net standing wave. On the other hand, if the interaction is attractive (i.e. $ \epsilon_p < 0$) then the former solution, corresponding to $\theta_\Delta=0$, is more stable, and there \underline{is} a standing wave. For a fixed $\hat{H}_{perturb}$, the magnitude of this energy gain depends strongly on the spatial (i.e. orbital) distribution of electron density, which corresponds to the spatial structure of $\psi_{1}$ and $\psi_{2}$. We note at this point that while we are taking a perturbative-type approach to describing the new states formed by an additional attractive interaction, this is not inconsistent with the famous result that the formation of a superconducting state is non-perturbative -- here, there is a discontinuous change in the solutions to this Hamiltonian when $\epsilon_p$ crosses 0. 

With this context we can further examine the variation in energy of a "single particle" superconducting wavefunction as a function of fixed $\theta_\Delta$ phase, and, by summing up over all equation \ref{eq:single_particle_pair} states, plot what the overall state look like in real-space. Choosing a Fermi arc ($f_a$) radius of $f_r = \frac{\pi}{8a}$, we can compute the phase-dependent energies, and the wavefunctions in real-space. Figure \ref{fig:Calculation}(a) shows that for some arbitrary $1/r$ attractive potential, the ground state will have $\theta_\Delta=0$. The wavefunction at select values of $\theta_\Delta$ can be seen in Figure \ref{fig:Calculation}(b-d), showing the standing wave characteristic of the superconducting state. We have thus established in our "single particle" Cooper pair picture that the superconducting state exists as a standing wave confined along a coherent path of orbitals between electrons of opposite momenta with a fixed phase difference.

\textbf{\begin{figure}
    \centering
    \includegraphics[height = 3.33in, width = 3.33in, keepaspectratio]{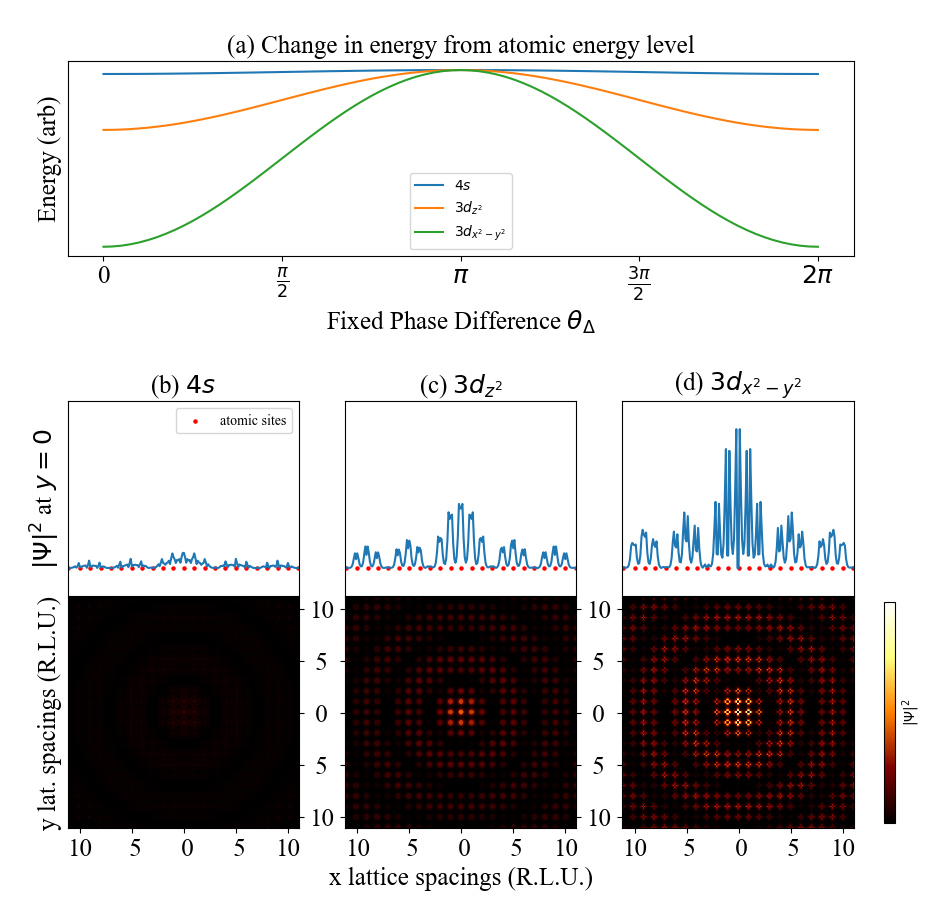}
    \caption{a) Energy of a "single particle" superconducting wavefunction as a function of phase difference calculated using a tight-binding model of 4$s$, 3$d_{z^2}$, and 3$d_{x^2 - y^2}$ orbitals. Probability density of finding such a state along atomic sites with $\theta_\Delta = 0$ using orbitals b) 4$s$ c) 3$d_{z^2}$, d) 3$d_{x^2 - y^2}$, where brighter coloration corresponds to higher probability.}
    \label{fig:Orbital_Comparison}
\end{figure}}

Up to this point we have only considered $s$ orbitals. However, it has long been known that the Fermi surface of cuprate superconductors is dominated by the copper $d_{x^2 - y^2}$ and oxygen $p_{x}$ and $p_{y}$ orbitals and, as a result, these orbital functions have become essential for describing the superconducting Hamiltonian  \cite{zhang_effective_1988}. In real-space, we can think of this particular tiling of orbitals on the square-net CuO$_{2}$ plane as permitting the formation of Cooper pairing. Therefore, it is important to explore more generally how the superconducting state changes in response to varying the configuration of the orbital tiling and the nature of the orbitals involved. 

Figure \ref{fig:Orbital_Comparison} shows the difference in energy for the "single particle" Cooper pair on tilings of 4$s$, 3$d_{z^2}$, and  3$d_{x^2 - y^2}$ orbitals with a uniformly chosen $1/r$ attractive interaction. Here lies the key observation of our toy model: we find that the orbital tilings have different depths of their energy minima, with $d_{x^2 - y^2}$ having the lowest energy! Note that this is still done without any explicit definition of the nature of the attractive potential. We have therefore demonstrated that, given a fixed phase difference between the electron wave functions, there is a particular value that minimizes the energy of the Cooper pair, and that the energy is further minimized upon appropriate orbital overlap and connectivity (i.e. the nature of the orbital tiling and the orbitals it is made from). 

Further, we know that in the non-superconducting state, electrons do not have this fixed momentum and phase relationship. Entering the superconducting state can now be visualized in real-space as the process by which nonsuperconducting delocalized electrons spontaneously transition from traveling along several orbital pathways with different phases to traveling only along a single orbital configuration, a transition which accounts for the breaking of gauge symmetry and the formation of a well-defined phase difference relation between electron pairs. Thus, our real-space model of the superconducting state is conceived as a standing wave of condensed Cooper pairs confined spatially, yet delocalized, along a particular orbital pathway. We refer to the orbital tiling configuration that defines this pathway as the Real-space Orbital Superconducting Pathway (ROSP). This real-space picture of conduction as the charge transfer across orbital pathways is analogous to the conduction pathways described in the MOF literature \cite{xie_electrically_2020}, and likewise, serves as a platform to engineer real-space architectures. 

We note that the ROSP model is consistent with the theoretical claim that superconductivity is orbital selective in nature, wherein the electrons of only specific orbitals are involved in Cooper pair formation \cite{arakawa_orbital-selective_2011}, which has been experimentally validated using scanning tunneling microscopy Bogoliubov quasiparticle interference imaging of the orbital nature \cite{sprau_discovery_2017,bi_orbital-selective_2022}. This ROSP picture is also consistent with the observation that orbital connectivity likely plays a significant role in superconducting properties. In the case of the cuprates, when the local coordination environments of the CuO$_{2}$ plane distort away from 90\textdegree - such that the $d_{x^2 - y^2}$ and $p_{x}$ and $p_{y}$ orbitals no longer overlap in a square net - superconductivity vanishes, as in Gd$_{2}$CuO$_{4}$ \cite{braden_structure_1994}. Moreover, minor changes in local coordination environment can result in significant property changes. For example, La$_{2-x}$Sr$_{x}$CuO$_{4}$ is an $\approx$ 40 K superconductor with an elongated octahedral coordination environment whose apical and equatorial Cu-O bond lengths are about 2.46 and 1.90 Å, respectively (ratio $\approx$ 1.29). HgBa$_{2}$CuO$_{4}$  on the other hand is a $\approx$ 90 K superconductor with a significantly more elongated octahedral environment where the apical and equatorial Cu-O bond lengths are $\approx$ 2.90 and 1.96 Å, respectively (ratio $\approx$ 1.48). Sakakibara et al. showed that the difference in T$_{c}$ could be understood due to the contribution of the $d_{z^2}$ orbital at the Fermi surface of La$_{2}$CuO$_{4}$ whereas the elongated environment in HgBa$_{2}$CuO$_{4}$ leads to a Fermi surface of strictly $d_{x^2 - y^2}$ character \cite{sakakibara_two-orbital_2010}. From the real-space coordination picture, it is clear how the observed differences in local bond distances between these two materials lead to significant changes in the orbital crystal field energy, and thus frontier orbital energy: as the ratio of the apical to equatorial bond lengths becomes closer to one, the $d_{z^2}$ orbital energy is brought closer to that of the $d_{x^2 - y^2}$, resulting in a lower T$_{c}$ or the loss of superconductivity altogether.

This is the key benefit of the ROSP model: even without microscopic knowledge of the attractive interaction, we can predict how variation in orbital arrangement affects the occurrence of superconductivity, and use this insight to design new superconductors. The ROSP, more generally, can be thought of as the subset of all possible orbital tilings which are compatible with superconductivity. It should be noted that orbital tilings are not truly ROSPs until the real-space orbital-selective nature of the Cooper pairs are experimentally verified. However, due to the novelty of these experimental measurements, and corresponding scarcity of them in the literature, computational and theoretical predictions will be used to assign ROSPs in the interim. 

Nonetheless, we can build intuition about what characteristics are necessary for a particular orbital tiling to be a ROSP. For a given orbital tiling to be conducive to a real-space Cooper pair standing wave, the electrons must delocalize, but for electron-electron interactions there is competition between a localized state and a delocalized superconducting state. The orbital tiling must therefore support a large kinetic energy gain from electron delocalization to stabilize superconductivity. This kinetic energy gain can be maximized when there is strong orbital overlap in multiple directions, allowing the attractive interaction to beat the tendency for localization. Qualitative orbital tilings that meet these electronic criteria serve as the starting point for our subsequent discussion of the design of novel superconductors.

\begin{figure*}
    \centering
    \includegraphics[height = 8in, width = 7in, keepaspectratio]{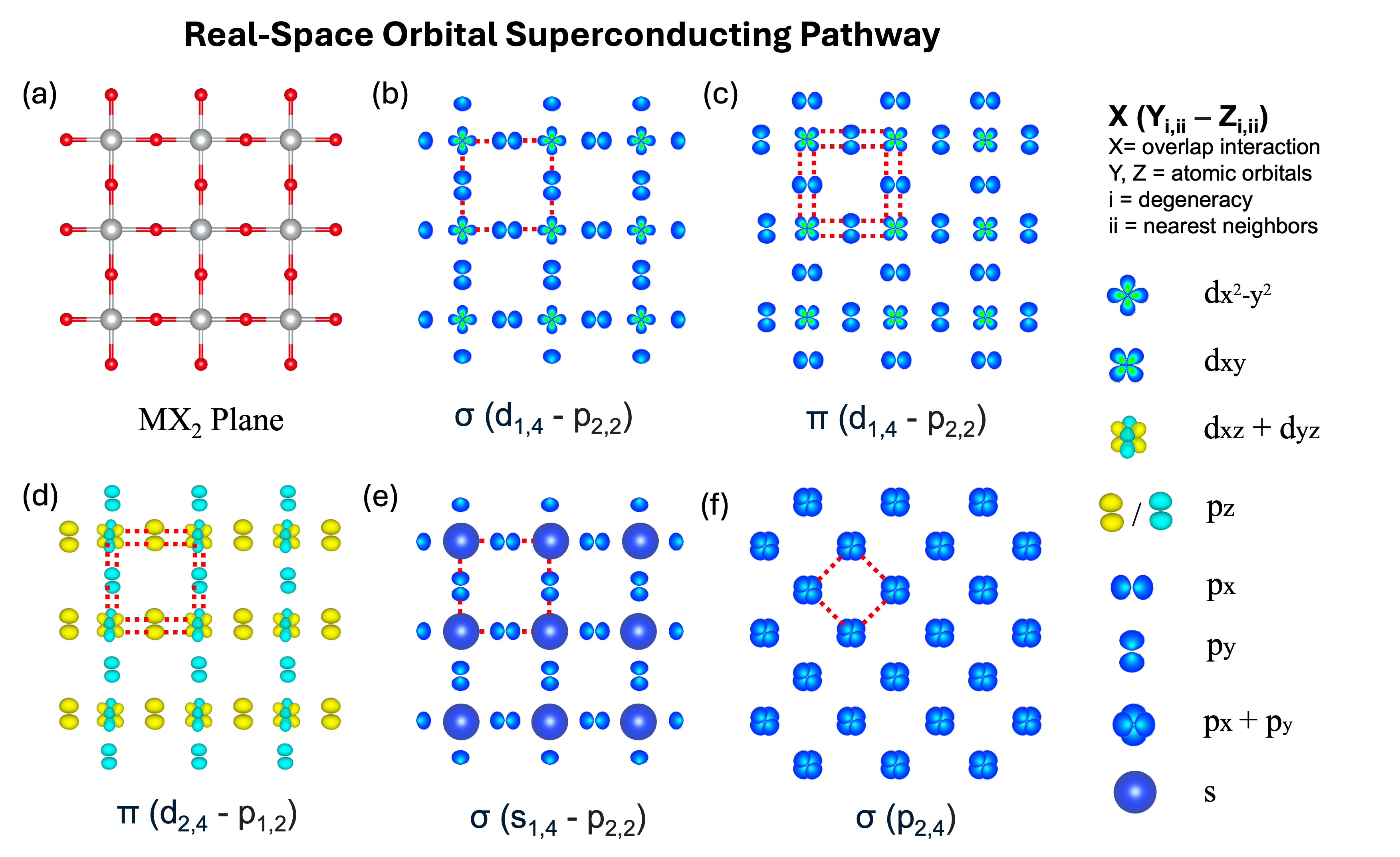}
    \caption{(a) MX$_{2}$ structural motif plane present in perovskites and Ruddlesden-Popper materials. (b-f) Schematic ROSPs from tilings of $s$, $p$, and/or $d$ orbitals. Red dashed lines correspond to orbital overlap, with single dash as $\sigma$ and double dash as $\pi$ overlap respectively. Notation is provided to assign each ROSP a unique identifier.}
    \label{fig:ROSP}
\end{figure*}
\subsection{Isolobality and Superconductor Classification}

Before we discuss our ROSP heuristics for designing superconductors, we must first describe how ROSPs will be classified. For experimentalists, the importance of electron count and structure for the discovery and classification of novel superconductors has long been established. As a result, it has become common practice to explore isostructural and isoelectronic analogs of known superconductors. For example, all cuprates were discovered by synthesizing materials with Cu in the same structural motifs and electron counts. Further, the recent discovery of superconductivity in thin film LaNiO$_{2}$ was motivated by a search for an isostructural and isoelectronic cuprate analog with Ni in the prototypical cuprate square-planar local coordination and  $d{^9}$ electronic configuration \cite{li_superconductivity_2019}. Another such example is the discovery of superconductivity in BaSbO$_{3}$, which is the isoelectronic and isostructural analog of BaBiO$_{3}$ \cite{kim_superconductivity_2022}. In these examples, the actual metal ion identity is only relevant insomuch as it can take both the same coordination environment and electron count as a related superconductor. However, isoelectronic and isostructural descriptors are not always successful in relating families of superconductors for discovery. For example, Ni in the recently discovered superconductor La$_{3}$Ni$_{2}$O$_{7}$ (under pressure) has a different oxidation state and coordination environment than in LaNiO$_{2}$ \cite{zhang_high-temperature_2024}, thus LaNiO$_{2}$ is neither isoelectronic nor isostructural La$_{3}$Ni$_{2}$O$_{7}$. 

This indicates the need for a new categorical mode to compare the familial similarity of superconducting materials. We propose that it is the nature of frontier orbital tiling -- agnostic to local structure and electron count -- that unifies materials into the same superconducting family. We therefore propose that superconductors should be classified by their ROSP, where two superconductors in the same family have the same ROSP and thus analogous tilings of frontier orbitals. We further call two materials that share a ROSP, even if they have different local structures and electron count, isolobal analogs. 

The root of this idea, that molecular fragments with distinct electron counts and coordination environments can have the same frontier orbitals and similar chemistry, was first pioneered by Roald Hoffmann. Hoffmann described isolobality in his Nobel lecture as when "the number, symmetry properties, approximate energy and shape of the frontier orbitals and the number of electrons in them are similar -- not identical, but similar" \cite{hoffmann_building_1982}. This isolobal analogy was developed to predict the bonding nature, reactivity, and structure of inorganic and organic molecular species by comparing the frontier orbitals. Recently, the isolobal concept has been applied for structural analysis in extended solids \cite{yannello_generality_2015}. As demonstrated by the previous nickelate example, the isolobal analogy is useful to expand beyond the classic thinking of isostructural and isoelectronic analogs of superconductors to unify systems with different local coordination environments and electron counts but analogous frontier orbital tiling in real-space. Thus, we extend the isolobal analogy beyond molecular fragment to the extended lattice of frontier orbital tiling more generally. It is through this isolobal analogy that ROSP families will be discussed. Specifically, two ROSPs will be considered isolobal if the approximate energy and shape of the frontier orbital tilings are analogous. 

With this context, we can now return to the comparison of superconductivity in La$_{3}$Ni$_{2}$O$_{7}$ and LaNiO$_{2}$. It is clear that LaNiO$_{2}$ is an isolobal analog of the cuprate superconductors, as it contains hybridized  $d_{x^2 - y^2}$ of Ni with  degenerate $p_{x}$ and $p_{y}$ of O within the NiO$_{2}$ plane \cite{anisimov_electronic_1999}. This further suggests that it is this orbital configuration which is essential for superconductivity in both LaNiO$_{2}$ and the cuprates, thus placing them in the same family. Like LaNiO$_{2}$, the La$_{3}$Ni$_{2}$O$_{7}$ band structure indicates density from Ni $d_{x^2 - y^2}$ and O $p_{x}$ and $p_{y}$ \cite{zhang_electronic_2023}. However, La$_{3}$Ni$_{2}$O$_{7}$ also has considerable contribution of the $d_{z^2}$ \cite{yang_orbital-dependent_2024}. We must therefore consider a few additional ROSP possibilities for La$_{3}$Ni$_{2}$O$_{7}$ beyond the $d_{x^2 - y^2}$ and $p_{x}$ and $p_{y}$ tiling of the cuprates /LaNiO$_{2}$. One possibility is that the O $p_{z}$ orbital and Ni $d_{z^2}$ may serve as a bridge between the $d_{x^2 - y^2}$ and degenerate $p_{x}$ and $p_{y}$ planes, forming a coherent interconnected bilayer cuprate ROSP. More interestingly, the $p_{x}$ and $p_{y}$ of O within the NiO$_{2}$ plane may overlap with the inner torus of Ni $d_{z^2}$ orbital under pressure, allowing for superconductivity to emerge in orbitals that would otherwise be too far apart. If this were the case, La$_{3}$Ni$_{2}$O$_{7}$ would have a distinct ROSP from the cuprates and LaNiO$_{2}$, serving as a new family prototype. We suspect that the apical oxygens are compressed slightly under pressure, favoring the $d_{x^2 - y^2}$ cuprate like ROSP. However, it remains to be seen whether or no La$_{3}$Ni$_{2}$O$_{7}$ and LaNiO$_{2}$ are isolobal analogs.  

This example shows how the ROSP model can be employed to classify and connect different superconductors. From this perspective, all members of a superconducting family are united by their shared ROSP, rather than a shared structural motif. Further, structurally similar superconductors with distinct superconducting states (ie. Sr$_{2}$RuO$_{4}$ and La$_{2}$CuO$_{4}$) can be differentiated according to their respective ROSP. Examples of ROSPs, both known and hypothetical, are shown in Figure \ref{fig:ROSP}. As the perovskite family of structures is known to host several classes of superconductors, the prototypical MX$_{2}$ plane [Figure \ref{fig:ROSP}(a)] is used as the structural building block from which interrelated ROSP schematics are generated. 

In order to more easily classify and represent these different layered 2D ROSPs we have developed simple notation identifiers, which take the form X(Y$_{i,ii}$-Z$_{i,ii}$). Here, X is the overlap interaction ($\sigma$ or $\pi$), Y is the atomic orbital ($s,p,d,f...$) of one atomic site, Z is a second atomic orbital site, $i$ is the orbital degeneracy, and $ii$ is the number of nearest neighbor orbitals from the corresponding orbital site. The dash separates unique orbital sites. This is repeated for all orbitals required to capture the tiling, with the orbital with the highest number of nearest neighbors listed first. For example, Figure \ref{fig:ROSP}(b) corresponds to the known essential orbital tiling layer of the cuprate superconductors, and now represents the family of ROSPs isolobal to this configuration. As previously described, this ROSP is built from $\sigma$  overlapping $d_{x^2-y^2}$ and degenerate $p_{x}$ and $p_{x}$ orbitals, each designated by red dashed lines in Figure \ref{fig:ROSP}(b). The shorthand notation can be written as $\sigma(d_{1,4}-p_{2,2})$, as the Cu site has a singly degenerate $d$-orbital with four nearest neighbors, and each O site is a doubly degenerate $p$ orbital with two nearest neighbors. Herein, $\sigma(d_{1,4}-p_{2,2})$ will refer to the name of the cuprate family ROSP which includes LaNiO$_{2}$, and we will also refer to this colloquially as the "cuprate ROSP."

Figure \ref{fig:ROSP}(c) illustrates a related ROSP $\pi(d_{1,4}-p_{2,2})$, a $\pi$ analog of the cuprates built from $d_{xy}$ and degenerate $p_{x}$ and $p_{x}$. There are no known superconductors that correspond to this ROSP, however it may eventually be realized by appropriate tuning of Sr$_{2}$IrO$_{4}$, a strongly pursued superconducting candidate. We will thus colloquially call this ROSP family the "iridate ROSP." Figure \ref{fig:ROSP}(d) corresponds to the ROSP $\pi(d_{2,4}-p_{1,2})$ built from degenerate $d_{xz}$ $d_{yz}$ orbitals with $p_{z}$, where the superconductor Sr$_{2}$RuO$_{4}$ is the prototypical family member. Figure \ref{fig:ROSP}(e) shows the ROSP $\sigma(s_{1,4}-p_{2,2})$ built from $s$ orbitals and degenerate $p_{x}$ and $p_{x}$ orbitals. The prototypical superconductor of this family is BaBiO$_{3}$, comprised of these BiO$_{2}$ planes. Finally, \ref{fig:ROSP}(f) corresponds to $\sigma(p_{2,4})$. This ROSP is unlikely to exist with oxygen in the MX$_{2}$ plane. However, as will be discussed in the ROSP Design section, this ROSP can likely be achieved more easily in the anti-cuprate M$_{2}$X structure. The orbital tilings of \ref{fig:ROSP}(b-d) are calculated using DFT to visualize the isosurfaces as shown in Figure \ref{fig:SI1.1}. In fact, the orbital tilings of \ref{fig:ROSP}(c) and \ref{fig:ROSP}(d) are present in the La$_{2}$CuO$_{4}$ band structure of Figure \ref{fig:SI1.1} deeper below the Fermi level. The calculation of these orbital tilings within the La$_{2}$CuO$_{4}$ structure suggests that achieving the desired ROSP is a matter of proper electronic filling within the structure. Thus, promising structural features like the MX$_{2}$ plane, can host multiple orbital tilings, and simply require precise electronic filling to align the corresponding ROSP states with the Fermi level. 

\subsection{Heuristics and Material Design Strategy}

This isolobal analogy will allow us to consider not only material phase spaces of known ROSPs, but provides a basis to explore novel ROSPs, the conditions for which are described in the following heuristics:  

\begin{enumerate}
    \item Combinations of frontier orbitals, or degenerate frontier orbitals, must overlap in a coherent pathway that tiles a plane (or a 3D version thereof).
    \item The overlap can be of sigma, pi, or delta character, and can be through-bond or through-space. 
    \item The tiling can be any combination of s, p, d, and f orbitals with its neighbors.
    \item Each orbital must occupy states at the Fermi level, where higher density of states are preferred. 
    \item The orbital tiling must have a net antiferromagnetic spin alignment and electronic correlation.  
\end{enumerate}

We now expand on the ideas behind these heuristics. The first condition for superconductivity is the presence of a coherent pathway for the Cooper pairs to travel. As mentioned previously, the orbital tiling must support a large kinetic energy gain from delocalization to prevent electronic localization and stabilize a standing wave.  Structurally, this means that there exists sufficient connectivity between local coordination environments such that frontier orbitals can tile a lattice, which is not feasible when the frontier orbitals are too far apart, as noted above. The nature of this overlap can be of overlapping sigma, pi, or delta character, and the pathway can be either through-bond or through-space (between bonding or nonbonding atoms respectively). If the orbital tiling is not at or near the Fermi level, there can be no charge transfer, as there need to be active electrons. From a molecular orbital and crystal field picture, the nature and coordination environment of ligands splits the energy levels of the orbitals, and the electron count determines which of these orbitals will be fully occupied. The energy of the Fermi level corresponds approximately to the band of the frontier orbital, especially when the frontier orbital is half filled \cite{hu_identifying_2016}. Moreover, the orbitals must be tiled with geometries such that the net exchange interaction between spins is antiferromagnetic, as antiferromagnetic alignment maximizes the chances that the interaction between electrons is attractive \cite{scalapino_common_2012}. For simplicity and clarity of presentation, we will only consider ROSP on a square net since the fourfold symmetry of d orbitals allows for straightforward overlapping along these lattices. Other lattices can also be used, but the orbital geometry is much less intuitive. 

\begin{figure}
    \centering
    \includegraphics[height = 3.33in, width = 3.33in, keepaspectratio]{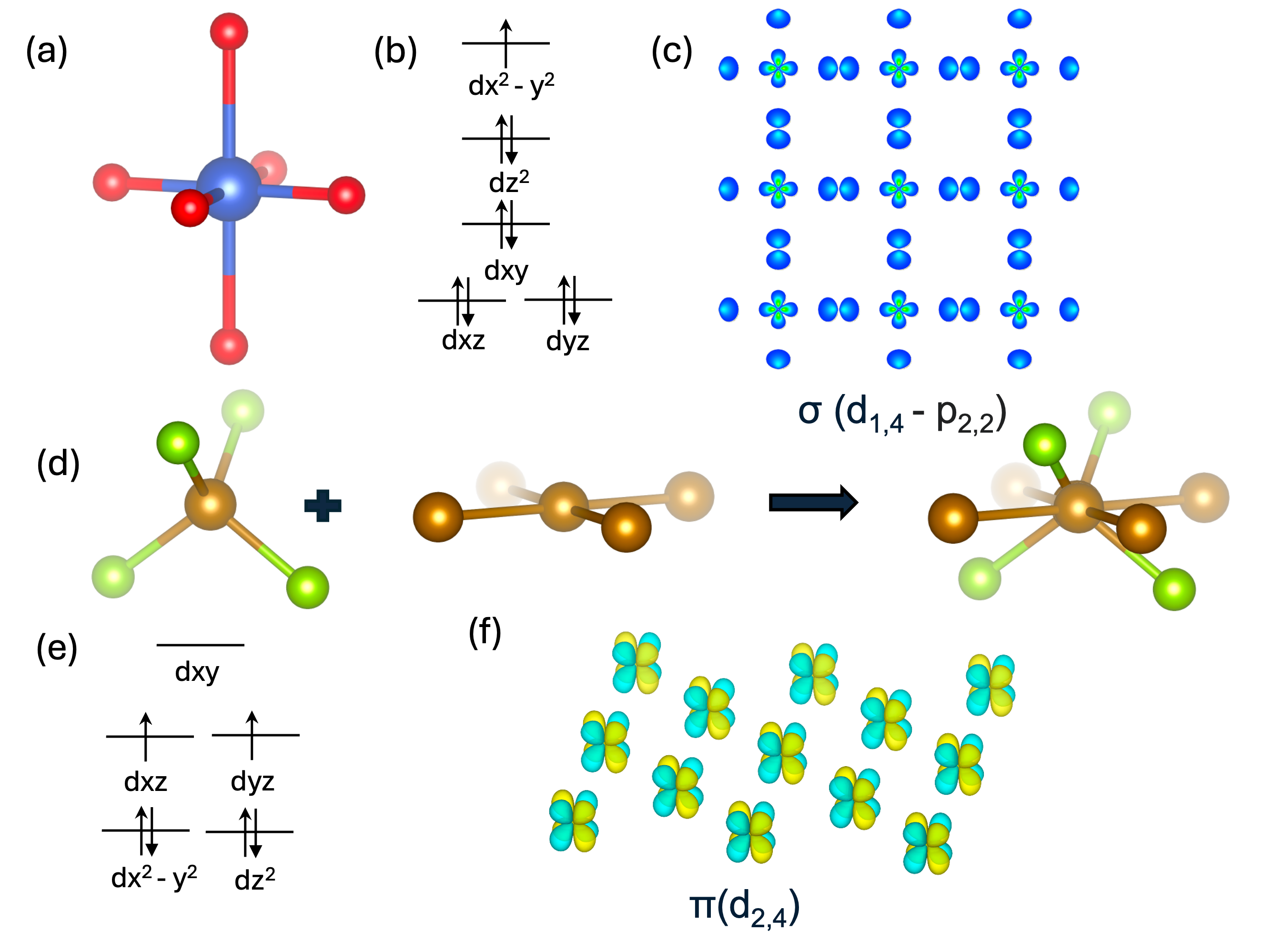}
    \caption{a) Elongated octahedra of copper within La$_{2}$CuO$_{4}$  b) crystal field splitting and electronic filling of $\mathrm{Cu}^{2+}$ in La$_{2}$CuO$_{4}$  c) ROSP in CuO$_{2}$ plane, $\sigma(d_{1,4}-p_{2,2})$ d) coordination environment of iron molecular fragment within FeSe e) crystal field splitting and electronic filling of $\mathrm{Fe}^{2+}$ in FeSe f) ROSP of the Fe plane, $\pi(d_{2,4})$}
    \label{fig:Design}
\end{figure}

\begin{figure*}
    \centering
    \includegraphics[height = 8in, width = 7in, keepaspectratio]{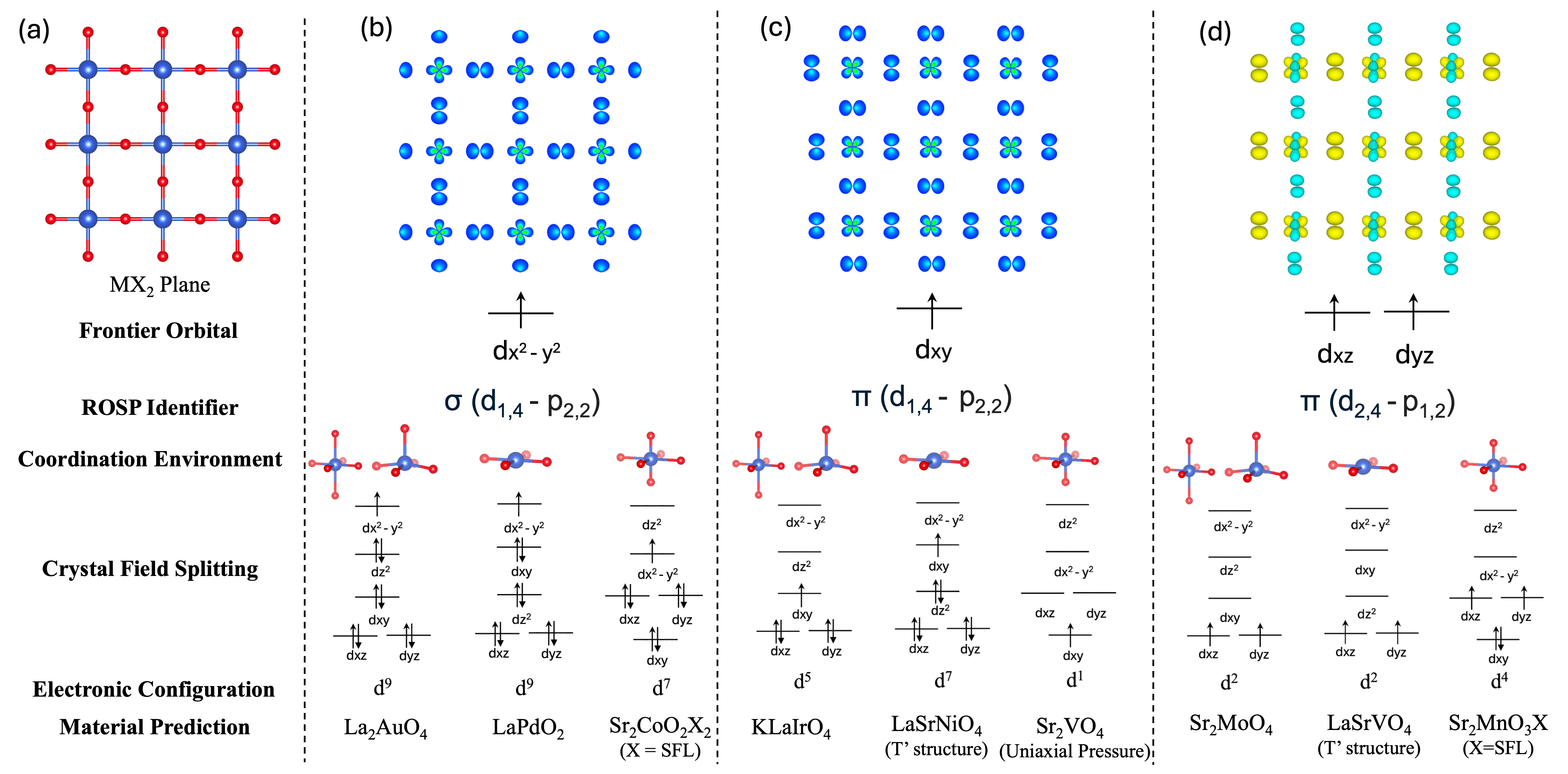}
    \caption{(a) MX$_{2}$ plane present in all Ruddlesden popper and perovskite derived structures. ROSP and corresponding coordination environment, crystal field splitting, and electronic configuration of predicted materials that would promote this ROSP. (b) ROSP family $\sigma(d_{1,4}-p_{2,2})$  from elongated octahedra d${^9}$, square planar d${^9}$, and compressed octahedra d${^7}$ low spin coordination environments. (c) ROSP family $\pi(d_{1,4}-p_{2,2})$ from elongated octahedra d${^5}$ low spin, square planar d${^7}$ low spin, and compressed octahedra d${^1}$ coordination environments. (d) ROSP family $\pi(d_{2,4}-p_{1,2})$ from elongated octahedra d${^2}$, square planar d${^2}$, and compressed octahedra d${^4}$ low spin coordination environments. SFL= Strong Field Ligand.}
    \label{fig:MX2}
\end{figure*}

The intuitive strategy for realizing targeted ROSPs in material systems is motivated by a molecular fragment approach to the relationship between crystal structure and electronic band structure. As the electrons involved in superconductivity are those at the Fermi energy, the crystal field splitting picture of orbitals is necessary insofar as it provides a framework to design the bands corresponding to the ROSP to be closest to the Fermi level. This approach allows for screening candidate materials by analyzing the local environment and computationally verifying if the expected ROSP band is present near the Fermi level. To illustrate the relationship of crystal field splitting and band structure, we will begin by analyzing the elongated octahedral coordination environment of copper in La$_{2}$CuO$_{4}$ as seen in Figure \ref{fig:Design}(a). In the case of the cuprates, degeneracy of the 3d orbitals is broken by the distorted octahedra, as shown in Figure \ref{fig:Design}(b). To find the frontier orbital in this manner, there must be sufficient charge separation between atoms to first assign a meaningful oxidation state and electron count. More generally, electronegativity differences between ions are a useful metric to determine the degree of charge separation, with ionicity increasing with larger electronegativity differences. In the cases of minimal electronegativity difference, it can be difficult to assign meaningful oxidation states, and thus frontier orbitals, \textit{a priori}. In the case of the cuprates, it is straightforward, as copper in this structure has a clear oxidation state of +2 and accordingly has a d${^9}$ electronic configuration. Based on the crystal field splitting diagram in Figure \ref{fig:Design}(b), the frontier orbital is the half-filled $d_{x^2 - y^2}$ orbital. Using DFT, an isosurface slice of the electronic probability density function can be calculated to visualize the orbital tiling at a given momentum space point. Figure \ref{fig:Design}(c) shows the ROSP $\sigma(d_{1,4}-p_{2,2})$ for La$_{2}$CuO$_{4}$ calculated at the high symmetry Z point nearest the Fermi level of the band structure. This orbital tiling is consistent with the expected  $d_{x^2 - y^2}$ on the copper site. 

A similar procedure can be repeated for the iron based superconductors. We take the example of FeSe, the simplest structure of the family. Here, iron takes a +2 d${^6}$ electronic configuration within a tetrahedral coordination environment of selenium. Figure \ref{fig:Design}(d) shows the local coordination environment of iron built from tetrahedra of selenium and nearest neighbor irons.  These local units are oriented together as an edge sharing square net, where each iron is bound to four neighboring iron sites. As in the example for La$_{2}$CuO$_{4}$, the frontier orbitals are singly occupied (half filled) but are now comprised of degenerate $d_{xz}$ and $d_{yz}$. These orbitals similarly tile the square net, but now through four nearest neighbor $\pi$ interactions (rather than sigma like the cuprates). Likewise, the band structure shows that the density of states at the Fermi level is dominated by iron $d_{xz}$ and $d_{yz}$ orbitals, and visualizing the electronic probability density function at the M point results in the expected orbital tiling as shown in (f). The band structure can be seen in  Figure \ref{fig:SI2.1}. 

\subsection{Material Predictions}

The MX$_{2}$ plane present in perovskite-derived structures is responsible for superconductivity in systems such as La$_{2}$CuO$_{4}$, Sr$_{2}$RuO$_{4}$, and LaNiO$_{2}$, and nicely illustrates the relationship between local coordination environment, structural connectivity and orbital configuration as shown in Figure \ref{fig:MX2}. Therefore, the MO2 structure serves as a simple platform to explore the isolobal analogy and ROSP. Figure \ref{fig:MX2}(b) shows the aforementioned ROSP $\sigma(d_{1,4}-p_{2,2})$, and below, describes the coordination environments and associated crystal field splitting/ electronic configurations that would result in this ROSP; material predictions for each case are also given. It should be noted that it is very uncommon to find materials whose frontier orbitals tile these configurations, likely due to the strong electronic instability. Superconductivity in the d${^9}$ cuprates has prompted a search for other isoelectronic and isostructural analogs, but d${^9}$ is a chemically unstable configuration in the other late transition metals, making the synthesis challenging through conventional solid-state means. Nevertheless, the recent discovery of superconductivity in topotactically reduced square planar nickel d${^9}$ LaNiO$_{2}$ thin films further motivates this discovery process, and likewise, we predict superconductivity in d${^9}$ LaPdO$_{2}$ and La$_{2}$AuO$_{4}$. $\mathrm{Pd}^{+}$  is a rather uncommon oxidation state for palladium, especially as an oxide. However, under certain low-temperature topotactic reducing conditions, it may be possible to deintercalate oxygen from a higher valent palladate, perhaps La$_{2}$Pd$_{2}$O$_{5}$, to isolate the desired LaPdO$_{2}$. Synthesizing La$_{2}$AuO$_{4}$ will also prove to be challenging. A possible synthetic route may involve an ion exchange from the more stable interrelated phase La$_{4}$LiAuO$_{8}$ (which can be thought of as a double n=1 ruddlesden popper perovskite). Gold naturally charge disproportionates into +1 and +3 oxidation states, and as the gold site is in a +3 state, starting with a Au (I) ion, perhaps AuCl, can be useful for exchanging with $\mathrm{Li}^{+}$ .  Beyond the isoelectronic d${^9}$ elongated octahedra or d${^9}$ square planar environment is the isolobal d${^7}$ low spin compressed octahedra. Square nets of compressed octahedra are exceedingly rare in nature, with the exception of the Ba$_{2}$ZnF$_{6}$ structure type \cite{von_schnering_kristallstrukturen_1967}. However, stabilization of this phase can be possible under sufficient uniaxial pressure or in the presence of strong field ligands in the apical site of the octahedra. Accordingly, the material Sr$_{2}$CoO$_{2}$(NO$_{2}$)$_{2}$ or perhaps Sr$_{2}$CoO$_{2}$(CN)$_{2}$ with strong field ligands in the apical sites would allow for the desired electronic configuration and orbital tiling. This synthesis, like the aforementioned aureate, may be possible under appropriate ion exchange conditions. Here, the precursor Sr$_{2}$CoO$_{2}$Cl$_{2}$ can perhaps be used in a low temperature topotactic metathesis reaction with NaX (X=CN$^-$, NO2$^-$) to exchange the strong field ligand for chlorine. We also predict superconductivity in materials like LaSrNiO$_{4}$, and LaSrVO$_{4}$ in the T' structure type. The synthesis of the T' structure may be difficult for these phases. Houchati et. al. found that La$_{2}$CuO$_{4}$ T' can be synthesized by low temperature reduction of the T phase to La$_{2}$CuO$_{3.5}$ using CaH$_{2}$ followed by a low temperature oxygenation \cite{houchati_phase_2021}. This suggests that both low temperature topotactic reductions and oxygenations may be necessary to achieve these phases, as well as the specialized design of new topotactic methods \cite{whoriskey_amalgams_2024}. Exploratory low temperature synthesis methods may also be needed to isolate these phases \cite{iwanicki_hydroflux-controlled_2024}.

The next family, $\pi(d_{1,4}-p_{2,2})$, is shown in Figure \ref{fig:MX2}(c). It has long been known that Sr$_{2}$IrO$_{4}$ holds promise for superconductivity due to its comparable Mott insulating behavior, one band Hubbard model and expected pairing symmetry to the cuprates. From an \textit{a priori} orbital tiling picture, we agree that the $\mathrm{Ir}^{4+}$  d${^5}$ system holds promise, but due to the distortions of the IrO$_{2}$ plane in Sr$_{2}$IrO$_{4}$ (rotations of the IrO$_{4}$ octahedra) and the similar distance between iridium and the apical and equitorial oxygens (non-isolated orbital energies), we instead believe that KLaIrO$_{4}$ would be a better representative material for this tiling. This material would take the KLaTiO$_{4}$ structure type, related to n=1 Ruddlesden popper, where the difference in ionic radii between $\mathrm{K}^{+}$  and $\mathrm{La}^{3+}$  causes an elongation of one apical oxygen, favoring the $d_{xy}$  frontier orbital with d${^5}$ configuration. However, due to the large spin-orbit coupling of iridium, the qualitative schematic of $d_{xy}$ orbitals may look different from the actual tiling because atomic orbitals form a poor basis in systems with high spin-orbit coupling.

\begin{figure*}
    \centering
    \includegraphics[height = 6in, width = 6in, keepaspectratio]{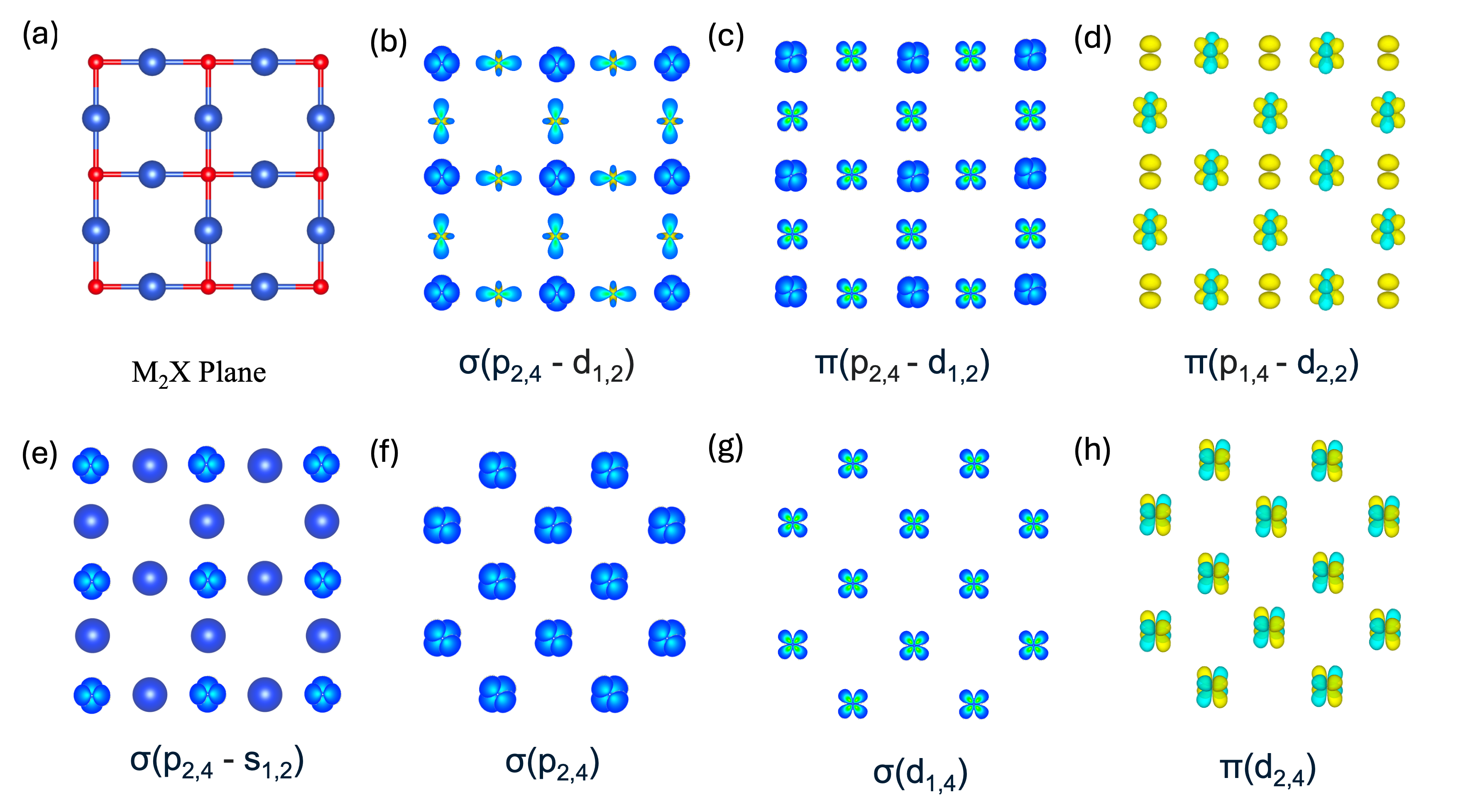}
    \caption{(a) M$_{2}$X layer. (b) 'anti-cuprate ROSP' family $\sigma(p_{2,4}-d_{1,2})$ built from  degenerate $p_{x}$ $p_{y}$ orbitals and  $d_{z^2}$. (c)'anti-iridate' ROSP family $\pi(p_{2,4}-d_{1,2})$ built from degenerate $p_{x}$ $p_{y}$ orbitals and $d_{xy}$. (d) 'anti-ruthenate ROSP' family $\pi(p_{1,4}-d_{2,2})$ built from $p_{z}$ and degenerate $d_{xz}$ $d_{yz}$ orbitals. (e) 'anti-bismuthate ROSP' family $\sigma(p_{2,4}-s_{1,2})$ built from degenerate $p_{x}$ $p_{y}$  with $s$-orbitals. (f) ROSP family $\sigma(p_{2,4})$ built from degenerate $p_{x}$ and $p_{y}$ orbitals. (g) ROSP family $\sigma(d_{1,4})$ built from degenerate $p_{x}$ and $p_{y}$ orbitals. (h) 'iron-based' ROSP family $\pi(d_{2,4})$ built from degenerate $d_{xz}$ and $d_{yz}$ orbitals}
    \label{fig:Anti_CuO2}
\end{figure*}

While we have outlined how the ROSP can be used to predict new superconductors, it needs to be emphasized that the existence of a ROSP is considered only a necessary, but not sufficient, condition for SC. The absence of superconductivity in systems which can be expected to host an ROSP must also be considered. For example, although superconductivity was discovered in doped LaFeAsO, it has not yet been observed in the isostructural, isoelectronic and isolobal LaRuAsO \cite{mcguire_structural_2012}. There are many reasons why this may be the case. For one, it is possible that the region of the bands containing the ROSP is shifted above/ below the Fermi level. Therefore, directed doping may be required to bring the fermi energy in line with the ROSP. We note that the material LaRu$_{2}$As$_{2}$ is a 7.8 K superconductor \cite{guo_superconductivity_2016}, suggesting that under sufficient electron doping, superconductivity can be achieved in the Ru analogs. The nature of the element, in this case iron or ruthenium, also affects the size of the relevant orbitals, which affects their overlap, or the degree of electronic correlation between sites. These differences may affect the doping conditions required for SC. 

\section{ROSP Design}

\begin{figure}
    \centering
    \includegraphics[height = 3.33in, width = 3.33in, keepaspectratio]{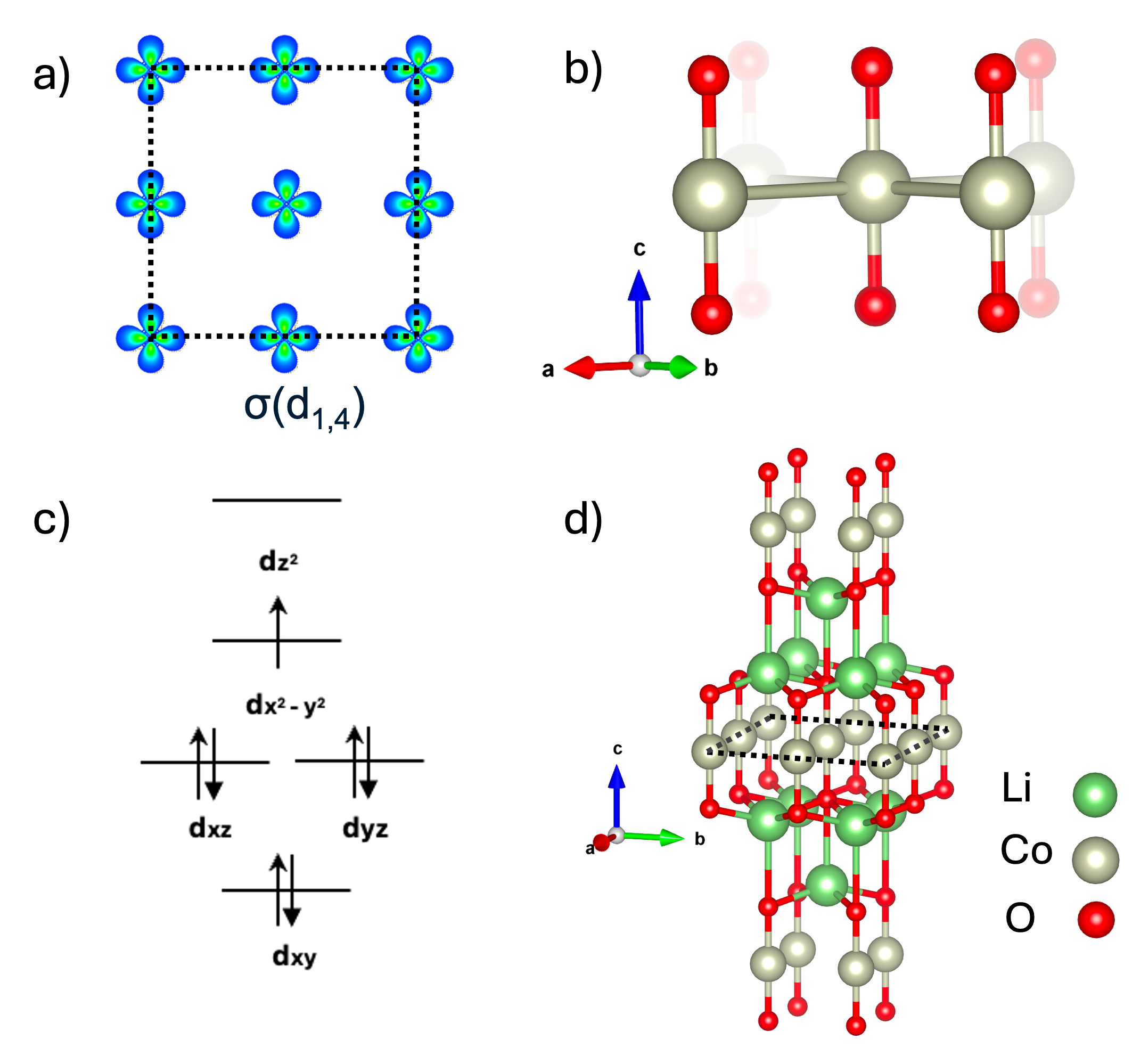}
    \caption{(a) ROSP $\sigma(d_{1,4})$ from $d_{x^2 - y^2}$  (b) molecular fragment of linear coordinated atoms (c) crystal field splitting for molecular fragment (d) proposed material Li$_{2}$CoO$_{2}$ and corresponding structure which satisfies these conditions}
    \label{fig:Ground_Up}
\end{figure}

The true utility of thinking in terms of ROSPs is that superconductors can now be designed retrosynthetically, by first qualitatively designing a ROSP, and then engineering a material whose valence electrons occupy said orbital tiling. For simplicity, we will begin with orbital tilings from known lattices and then show an exercise of a material design from the ground up ROSP case. 

Much like the MX$_{2}$ plane discussed previously, the anticuprate M$_{2}$X plane, shown in Figure \ref{fig:Anti_CuO2}(a), allows for a variety of tilings with different frontier orbitals; however, this structure type is much less explored. A brief overview of known materials may be useful to motivate possible structure types and chemical phases spaces containing this plane. For example, the material BaTi$_{2}$Sb$_{2}$O is a known superconductor containing the anticuprate plane. \cite{yajima_superconductivity_2012}. Other materials containing this M$_{2}$X plane include RbV$_{2}$Te$_{2}$O, CsV$_{2}$S$_{2}$O, Na$_{2}$Fe$_{2}$S$_{2}$O, La$_{2}$O$_{3}$Co$_{2}$Se$_{2}$ and ThCr$_{2}$Si$_{2}$C, with a Cr$_{2}$C layer \cite{xiao_thcr2_2024,valldor_badmetallayered_2016,ablimit_weak_2018,he_synthesis_2011}. ThCr$_{2}$Si$_{2}$C is a notable structure as the carbon intercalation changes the conventional ThCr$_{2}$Si$_{2}$ structure from that of the iron based superconductors with a square net of edge sharing CrSi$_{4}$ tetrahedra, to the anticuprate layered structure of square net face sharing CrSi$_{4}$C$_{2}$ octahedra. Further, superconductivity was discovered in ThMo$_{2}$Si$_{2}$C \cite{liu_superconductivity_2021}. As ThCr$_{2}$Si$_{2}$ is an enormous structure type and can be converted to the anticuprate structure through intercalation, this suggests an enormous untapped phase space of anticuprate layered materials. 

Accordingly, we propose several ROSPs inspired from this lattice, and we suspect that some are isolobal to other known superconductors. To avoid limiting the material structure types and phase spaces, we do not specify coordination environments or crystal field splittings that would result in these frontier orbital tilings. 
Figure \ref{fig:Anti_CuO2}(b) shows the family $\sigma(p_{2,4}-d_{1,2})$. We note that this ROSP is similar to the 'cuprate ROSP' $\sigma(d_{1,4}-p_{2,2})$, with the same degeneracy on the $d$ and $p$ orbitals respectively, and with the expected inverted nearest neighbor count. Accordingly, we colloquially call this the 'anti-cuprate ROSP'. Qualitatively, the degenerate $p_{x}$ and $p_{y}$ resembles the symmetry of the $d_{x^2-y2}$ orbital, and the in-plane $d_{z^2}$ orbitals resembles the lobes of the $p_{x}$/ $p_{y}$ orbital. A similar 'inverted' situation is observed for ROSP family $\pi(p_{2,4}-d_{1,2})$  in Figure \ref{fig:Anti_CuO2}(c), where it is the $\pi$ analog of the anti-cuprate ROSP (remember, the iridate is the $\pi$ analog of the cuprate ROSP) and accordingly, is colloquially the 'anti-iridate ROSP'. This trend continues for the family $\pi(p_{1,4}-d_{2,2})$ in Figure \ref{fig:Anti_CuO2}(d), which can be seen as the 'anti-Ruthenate ROSP.' The last of these 'anti-ROSPs' is seen in Figure \ref{fig:Anti_CuO2}(e), $\sigma(p_{2,4}-s_{1,2})$, or the 'anti-bismuthate.' The subsequent ROSPs are built from one orbital type, and thus have clear isolobal analogs from other structure types. Figure \ref{fig:Anti_CuO2}(f) shows the family $\sigma(p_{2,4})$, which is isolobal to the orbital tiling of Figure\ref{fig:ROSP}(f) and may have been experimentally realized in the arsenic square net of superconducting Li$_{0.6}$Sn$_{2}$As$_{2}$ \cite{wang_superconductivity_2023}. Figure \ref{fig:Anti_CuO2}(g) depicts the ROSP family $\sigma(d_{1,4})$. One can imagine this ROSP in the iron-based structure type from a nominal $d^9$ configuration of $d_{xy}$ orbital tiling, and thus realization of Cu$^{2+}$ or Ni$^{1+}$/ Pd$^{1+}$ $d^9$ iron-based structure type would presumably be isolobal. Finally, Figure \ref{fig:Anti_CuO2}(h), shows the family $\pi(d_{2,4})$, which is characteristic of the iron based superconductors, suggesting the anticuprate structure type may host many new isolobal analogs to the high-T$_{c}$ iron based superconductors. 

With this basis, one can now envision designing superconductors entirely from the ground up. We will begin by targeting a simple ROSP, $\sigma(d_{1,4})$, much like that calculated in Figure \ref{fig:Orbital_Comparison}, the anticuprate Figure \ref{fig:Anti_CuO2}(b), and the $d^9$ iron based structure examples discussed previously. Next, we select a molecular fragment and coordination environment that can tile a plane with its neighbors. Here we will use a linear coordination environment as shown in Figure \ref{fig:Ground_Up}(b), whose tiling is reminiscent of the delafossite structure. We then half-fill the desired frontier orbital using the corresponding number of electrons (low spin) from the derived crystal field. In this case, it is d${^7}$ as shown in Figure \ref{fig:Ground_Up}(c). We now need an element whose oxidation states allows for access to this frontier orbital: we will select $\mathrm{Co}^{2+}$. Finally, we require additional structural scaffolding to serve as charge reservoirs and/ or spacers between the ROSP layer. In this case, square pyramidal $\mathrm{Li}^{+}$ is used. The resulting structure takes the Na$_{2}$HgO$_{2}$ structure type. The synthesis and experimental realization of this material would be very challenging, as $\mathrm{Co}^{2+}$ is not stable in a linear coordination environment. Further, the linear Co-O bond distances would have to be sufficiently small to favor a low spin environment, which make require additional pressure. However, the primary result from this simple exercise is to illustrate how superconductors can be designed \textit{a priori} from orbital tiling. 

Based on this exercise, the logical next generation of superconductor discovery may consist of using generative computational models to construct stable structures containing all of these parameters concurrently guided by ROSP heuristics. Perhaps this can be done by constructing variations of coordination environments around a central atom (capable of tiling 2D space), with the crystal field calculated and filled with electrons accordingly. Provided that the computation model possess sufficient chemical knowledge (e.g. which elements take which oxidation states and coordination environments), this would allow for the promising generation of candidate superconductors. Moreover, the resulting generated structure will not be limited to known structural motifs, or related versions thereof if assembled purely locally and set to minimize the energy of the structure.

\section{Conclusion}

Here we propose the idea of the Real-space Orbital Superconducting Pathway (ROSP) which promotes the visualization of superconductivity along discrete tilings of orbitals. The ROSP model is useful as a novel approach for designing superconductors and allows for more simple classification of superconductors into families. This was informed by theoretical models of electron-electron attractive interaction on a lattice. We describe a design strategy motivated by local coordination environments and electron counting as a method for realizing targeted ROSPs. We further employ the isolobal analogy to compare La$_{3}$Ni$_{2}$O$_{7}$ and LaNiO$_{2}$ which are neither isostructural nor isoelectronic analogs, but may be related through a shared orbital tiling. We propose several candidate superconducting materials using ROSPs. 

We have shown how the ROSP model allows for the categorization of existing superconductors and the prediction of new ones based on an intuitive, real-space picture. However, ROSPs can also be useful to understand phenomenological properties of superconductors more generally. For example, it is well known that temperature is antithetical to superconductivity. In real-space, as temperature increases, the central atom has more energy to displace from its neighboring ligands, causing fluctuations to local orbital overlap. Moreover, the average bond lengths of the local coordination environment tend to increase, further minimizing orbital overlap and destroying the ROSP. On the other hand, high pressure is known to aid in the onset of superconductivity. In real-space, this serves the opposite function of temperature, wherein the atoms are pushed closer together, increasing the orbital overlap between neighboring sites, maintaining or even inducing a coherent superconducting pathway.

We believe that the ROSP model may serve as a general classification for superconductivity which goes beyond any specific family or pairing mechanism. While the above discussion is in the context of unconventional superconductors, it appears completely possible that the ROSP viewpoint could be employed for rationalizing and discovering conventional superconductors as well. To further elaborate, BCS theory showed that any arbitrary attractive potential between electrons results in the formation of Cooper pairs lowers the overall energy of the system through the formation of a condensate phase due to their bosonic nature  \cite{Bardeen1957}. In the case of conventional (low-$T_c$) superconductors, experiments with isotopic substitution demonstrated that the attraction is due to electron-phonon interactions \cite{maxwell_isotope_1950}. Unconventional superconductors also possess an attractive potential that facilitates electron pairing, but while this interaction is known to be distinct from the electron-phonon interaction in conventional superconductors, the exact mechanism remains unknown. In either case, whether conventional or unconventional pairing, the energy of the system remains dependent on the spatial variation of the electronic wavefunctions, and thus the nature of the orbital tiling. 

We note that significant chemical and synthetic challenges must be overcome to experimentally realize candidate materials. We urge experimentalists to explore the anticuprate structural plane, which holds promise as the logical successor to the cuprate plane. Finally, we walk through the design of a superconductor candidate from the ground up based on our orbital tiling approach. 

\section{Acknowledgments}

We thank Professor Yi Li for reviewing an early version of the manuscript. We thank the Johns Hopkins University for support of this work. RE acknowledges support from the David and Lucile Packard foundation through the FiGURE program. EZ acknowledges support from the Sweeney Postdoctoral Fellowship.

\appendix*
\counterwithin{figure}{section}
\renewcommand{\thefigure}{A\arabic{figure}}
\section{DFT methods}
 Band structure calculations were performed using Elk, an all-electron full-potential linearised augmented-plane wave code, with a $k$-point mesh of 10x10x10. The Perdew-Burke-Ernzerhof (PBE) parametrization of the generalized gradient approximation exchange-correlation functional was used. The cutoff energy used for augmented plane wave functions was 150 eV; the convergence criteria for root-mean-square change in Kohn-Sham potential and absolute change in total energy were at least 10$^{-6}$ eV and 10$^{-5}$ eV respectively. 
\begin{figure*}
    \centering
    \includegraphics[height = 5in, width = 5in, keepaspectratio]{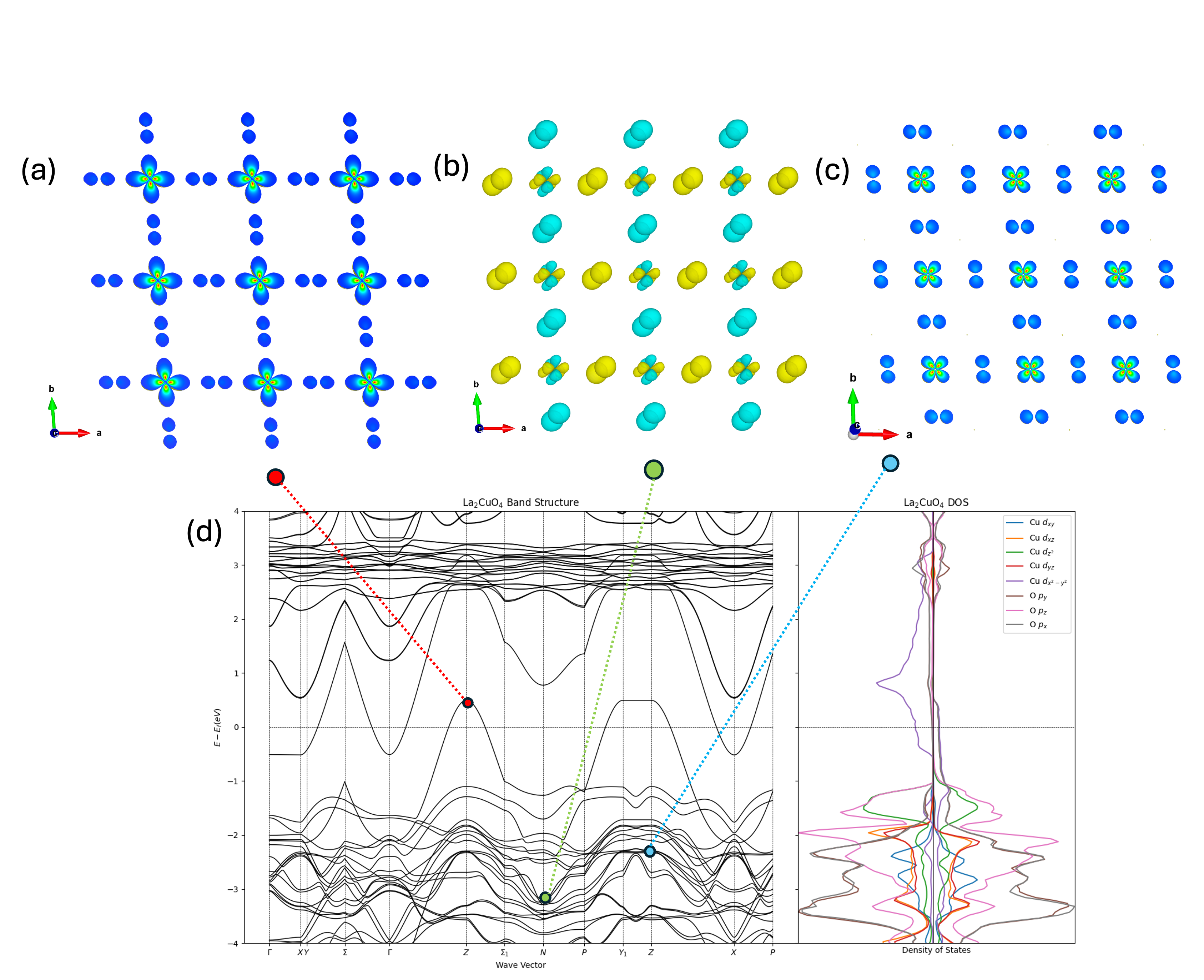}
    \caption{Orbital isosurface for La$_{2}$CuO$_{4}$ calculated from symmetry points a) Z b) N c) Z where dot color corresponds to the point on band structure where real-space orbitals were calculated. d) calculated band structure for La$_{2}$CuO$_{4}$ and corresponding DOS.}
    \label{fig:SI1.1}
\end{figure*}

\begin{figure*}
    \centering
    \includegraphics[height = 5in, width = 5in, keepaspectratio]{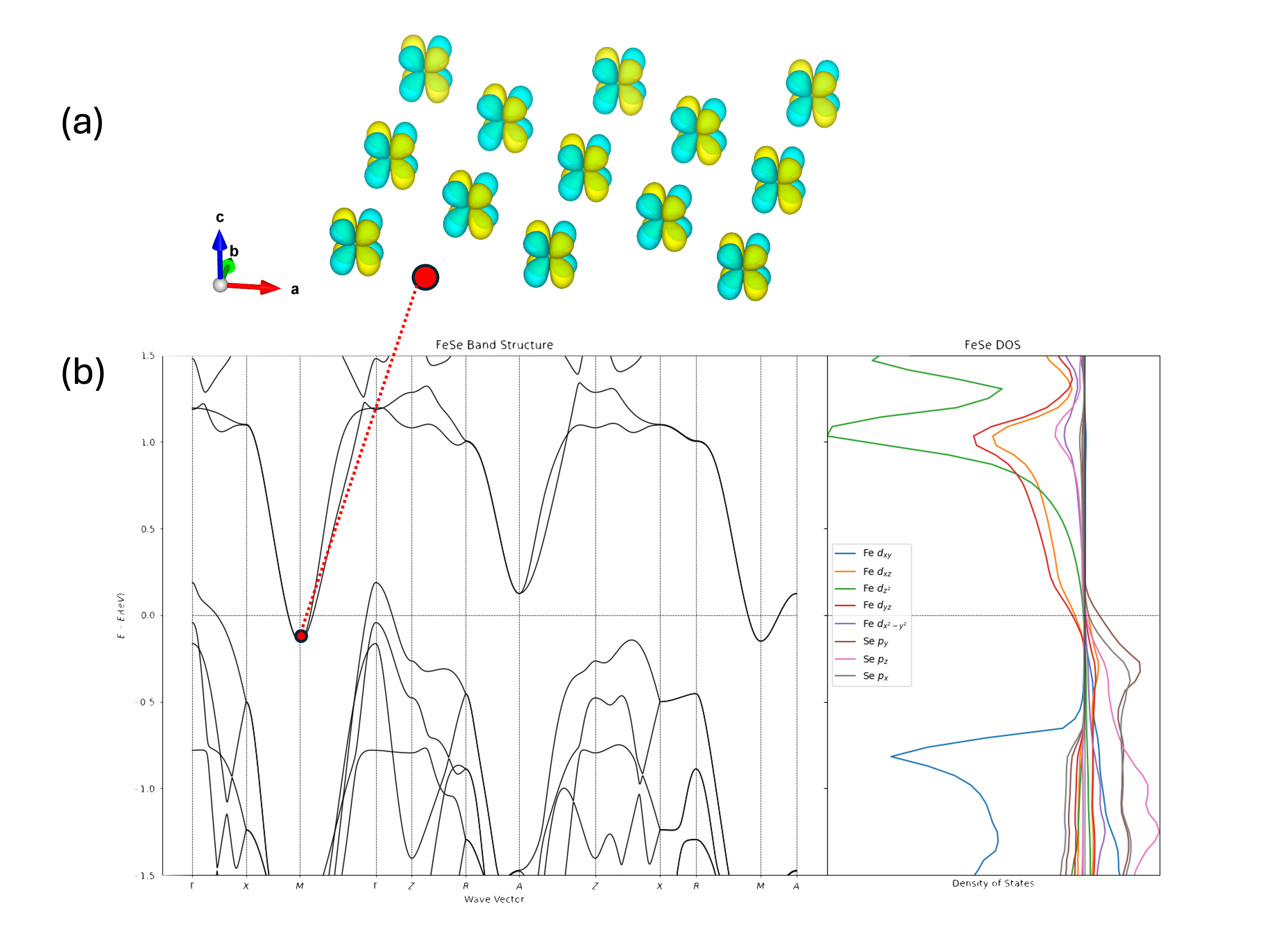}
    \caption{a) Orbital isosurface for FeSe calculated from M point. Red dot corresponds to the point on band structure where real-space orbitals were calculated. b) calculated band structure for FeSe and corresponding DOS.}
    \label{fig:SI2.1}
    
 \end{figure*}
 
\bibliography{main.bib}
\bibliographystyle{achemso}

\end{document}